\documentclass[10pt,letterpaper]{article}

\usepackage{graphicx}
\usepackage{url}
\usepackage{pslatex}
\usepackage{amsmath,amssymb}
\usepackage{algorithm}
\usepackage{algorithmic}
\usepackage[usenames,dvipsnames]{color}
\usepackage{fancyhdr}
\usepackage{subfig}
\usepackage{multicol}
\usepackage{multirow}
\usepackage{caption}
\usepackage{wrapfig}
\usepackage{booktabs}
\usepackage{multirow}
\usepackage{xspace}
\newcommand{\kevin}[1]{}
\newcommand{\lavanya}[1]{}

\newcommand{\MODIFIEDISCA}[1]{#1}
\newcommand{\MODIFIED}[1]{#1}
\newcommand{\ADD}[1]{#1}

\newcommand{\TMP}[1]{#1}

\newcommand{\ca}[0]{\texttt{CPU-A}\xspace}
\newcommand{\cb}[0]{\texttt{CPU-B}\xspace}
\newcommand{\hwa}[0]{\texttt{HWA}\xspace}

\newcommand{\TODO}[1]{}

\newcommand{\ldphwas}[0]{LDP-HWAs\xspace}

\newcommand{\sdphwas}[0]{SDP-HWAs\xspace}
\newcommand{\sdphwa}[0]{SDP-HWA\xspace}

\newif\ifSQUEEZE
\SQUEEZEtrue
\newif\ifOPTION
\OPTIONtrue
\usepackage[top=1in,bottom=1in,left=1in,right=1in]{geometry}

\usepackage[pagebackref=true,breaklinks=true,letterpaper=true,colorlinks,bookmarks=false]{hyperref}

\newcommand{\ignore}[1]{}

\usepackage[normalem]{ulem}

\fancypagestyle{plain}{
\fancyhf{}  %clear all header and footer fields
\chead{SAFARI Technical Report No. 2015-003 (March 18, 2015)}
\cfoot{\thepage}
}

\begin{document}

\title{\Large{SQUASH: Simple QoS-Aware High-Performance Memory Scheduler\\for Heterogeneous Systems with Hardware Accelerators}}
%HiRD: A Low-Complexity, Energy-Efficient \\ Hierarchical Ring Interconnect}

%\author{
%\begin{tabular}[t]{ccc}
%  Hiroyuki Usui & Lavanya Subramanian & Kevin Chang \\
%  \texttt{husui@andrew.cmu.edu} & \texttt{lsubrama@andrew.cmu.edu} & \texttt{kevincha@andrew.cmu.edu} \\
%  & Onur Mutlu & \\
%  & \texttt{onur@cmu.edu} & \\
%\multicolumn{3}{c}{} \\
%%\multicolumn{3}{c}{Computer Architecture Lab (CALCM)} \\
%\multicolumn{3}{c}{Carnegie Mellon University} \\
%%\multicolumn{3}{c}{} \\
%%\multicolumn{3}{c}{$\dagger$Massachusetts Institute of Technology} \\
%%\multicolumn{3}{c}{$\S$University of Michigan} \\
%\multicolumn{3}{c}{} \\
%\multicolumn{3}{c}{SAFARI Technical Report No. 2015-XXX}
%\end{tabular}
%}
\author{
\begin{tabular}[t]{cc}
%  Hiroyuki Usui & Lavanya Subramanian & Kevin Chang \\
  Hiroyuki Usui & Lavanya Subramanian \\
  \texttt{husui@andrew.cmu.edu} & \texttt{lsubrama@andrew.cmu.edu} \\
  Kevin Chang & Onur Mutlu \\
  \texttt{kevincha@andrew.cmu.edu} & \texttt{onur@cmu.edu} \\
\multicolumn{2}{c}{} \\
%\multicolumn{3}{c}{Computer Architecture Lab (CALCM)} \\
\multicolumn{2}{c}{Carnegie Mellon University} \\
%\multicolumn{3}{c}{} \\
%\multicolumn{3}{c}{$\dagger$Massachusetts Institute of Technology} \\
%\multicolumn{3}{c}{$\S$University of Michigan} \\
\multicolumn{2}{c}{} \\
\multicolumn{2}{c}{SAFARI Technical Report No. 2015-003}
\end{tabular}
}

\date{March 18, 2015}

\maketitle

\begin{abstract}

Modern SoCs integrate multiple CPU cores and Hardware Accelerators
(HWAs) that share the same main memory system, causing interference
among memory requests from different agents. The result of this
interference, if not controlled well, is missed deadlines for HWAs and
low CPU performance.
%Few previous works have tackled the problem of
%meeting HWAs' deadlines while achieving high CPU performance. 
State-of-the-art mechanisms designed for CPU-GPU systems strive to
meet a target frame rate for GPUs by prioritizing the GPU close to the
time when it has to complete a frame. We observe two major problems
when such an approach is adapted to a heterogeneous CPU-HWA
system. First, HWAs miss deadlines because they are prioritized
\emph{only when} they are too close to their deadlines. Second, such an
approach does \emph{not} consider the diverse memory access
characteristics of different applications running on CPUs and HWAs,
leading to low performance for latency-sensitive CPU applications
and deadline misses for some HWAs, including GPUs.

%Current SoCs are heterogeneous architectures that integrate CPU cores and
%Hardware Accelerators (HWAs). CPU cores and HWAs share the main memory causing
%interference between their requests. The result of this interference is missed
%deadlines for the HWAs and low CPU performance. Few previous works have tackled
%the problem of mitigating interference between CPU cores and HWAs, with the goal
%of enabling HWAs to meet their deadlines, while also improving CPU performance.
%Previous work has designed memory management techniques for CPU-GPU systems,
%where the GPU is given priority when it needs to complete a frame, in order to
%meet a target frame rate. When we adapt such a technique to manage interference
%between CPU cores and HWAs, we observe two major problems. First, HWAs are
%prioritized only when close to a deadline, causing deadline misses. Second, such
%a scheme does not take into account the memory access characteristics of
%different applications running on CPU cores and HWAs.

In this paper, we propose a Simple QUality of service Aware memory
Scheduler for Heterogeneous systems (SQUASH), that overcomes these
problems using three key ideas, with the goal of meeting HWAs'
deadlines while providing high CPU performance. First, SQUASH
prioritizes a HWA when it is not on track to meet its deadline 
\emph{any time} during a deadline period, instead of prioritizing it only when
close to a deadline.  Second, SQUASH prioritizes HWAs over
memory-intensive CPU applications based on the observation that
memory-intensive applications' performance is not sensitive to memory
latency.
%SQUASH exploits the observation that memory-intensive CPU cores' performance is
%not affected significantly when their requests are delayed and prioritizes HWAs
%over memory-intensive CPU applications, even when HWAs are making sufficient
%progress.
%Third, SQUASH observes that a progress-based priority scheme could cause
%deadline misses for short-deadline-period HWAs that generate few memory
%requests. Therefore, it gives them highest priority for a short burst of time
%close to their deadline, by means of estimating their worst-case memory latency.
Third, SQUASH treats short-deadline HWAs differently as they are more
likely to miss their deadlines and schedules their requests based on
worst-case memory access time estimates.
%observes that a progress-based priority scheme could cause
%deadline misses for short-deadline-period HWAs that generate few memory
%requests. Therefore, it gives them highest priority for a short burst of time
%close to their deadline, by means of estimating their worst-case memory latency.

Extensive evaluations across a wide variety of different workloads and systems
%show that SQUASH achieves 14.8\% better CPU performance than the best previous
%show that SQUASH achieves 13.9\% better CPU performance than the best previous
%show that SQUASH achieves 10.1\% better CPU performance than the best previous
show that SQUASH achieves significantly better CPU performance than the best previous
scheduler while always meeting the deadlines for all HWAs, including GPUs, thereby
largely improving frame rates.

%Current SoCs are heterogeneous architectures that integrate CPUs and Hardware Accelerators (HWAs).
%In the SoCs, CPUs and HWAs share the same main memory subsystem and interfere with each other in main memory.
%The interference on the main memory makes HWA miss the deadlines and slowdown CPUs. Memory scheduler in such
%heterogeneous architecture is required to improve CPU performance while meeting HWAs' deadline.
%Challenge in the memory scheduler is that different HWAs have different memory access characteristics and
%different deadlines, which state of the art memory schedulers do not smoothly handle. Memory-intensive and
%long-deadline HWAs significantly degrade system performance when they become higher priority than CPUs.
%Short-deadline HWAs somtimes miss their deadlines despite high priority.
%
%In this paper, we propose our Simple QUarity of service Aware memory Scheduler for Heterogeneous systems (SQUASH) using
%two key ideas. First, SQUASH prevents a HWA from becoming high priority by prioritize the HWA over a memory-intensive
%core even when the HWA is making good progress. Second, short-deadline HWAs are prioritized over long-deadline HWAs
%and CPUs based on the worst-case memory access time estimation.
%
%We evaluate SQUASH across a wide variety of differnt workloads and system configurations. Our evaluation shows
%the SQUASH achieves from 7\% to 21\% better system performance compared with state of the art memory scheduler and
%100\% deadline met ratio and highest frame rates for all HWAs.

\end{abstract}

\section{Introduction}

Today's SoCs are heterogeneous architectures that integrate
hardware accelerators (HWAs) and CPUs. Special-purpose hardware
accelerators are widely used in SoCs, along with general-purpose
CPU cores, because of their ability to perform specific operations
in a fast and energy-efficient manner. For example, CPU cores and
Graphics Processing Units (GPUs) are often integrated in smart
phone SoCs~\cite{snapdragon}. Hard-wired
HWAs are implemented in a very wide range of SoCs~\cite{visconti2}, including
smart phones.
% Citations are needed for current mobile SoCs besides Visconti2

%The HWAs are designed
%so that target performance (e.g., frame rate) is satisfied in the design phase. However, HWAs share the memory subsytem with CPUs and HWAs cannot
%satisfy target performance due to interference form CPUs. Therefore, it is important and challenging to achieve
%high system performance while keeping HWAs' performance.

In most such SoCs, HWAs share the main memory with CPU cores. Main
memory is a heavily contended resource between multiple agents and
a critical bottleneck in such systems~\cite{snapdragon,visconti2}.
This becomes even more of a
problem since HWAs need to meet \emph{deadlines}. Therefore, it is
important to manage the main memory such that HWAs meet deadlines
while CPUs achieve high performance.

Several previous works have explored application-aware memory
%request scheduling\cite{stfm,fair-queueing-memory,parbs,atlas,tcm},
request scheduling in CPU-only multicore systems~\cite{stfm,fqm,parbs,atlas,tcm,blist}.
The basic idea is to reorder requests from different CPU cores
to achieve high performance and
fairness. However, there have been few previous works that have tackled the problem of
main memory management in heterogeneous systems consisting of CPUs and HWAs, with the
dual goals of 1) meeting HWAs' deadlines while 2) achieving high CPU performance.

The closest works tackle the problem of memory management in specifically CPU-GPU systems
(e.g.,~\cite{schedulingCPUGPU,sms}). In particular, the state-of-the-art memory
scheduling scheme for CPU-GPU systems~\cite{schedulingCPUGPU} strives to meet a
target frame rate for the GPU while achieving high CPU performance. Its key idea
is to prioritize the GPU over the CPU cores \emph{only close to the deadline}
when the GPU has to finish a frame. The GPU is either deprioritized or given
the same priority as the CPU cores at other times. This scheme does not 
consider different HWAs than GPUs.

We adapted this state-of-the-art scheme~\cite{schedulingCPUGPU} to a
heterogeneous system with CPUs and various HWAs, and observed that such an approach when
used in a CPU-HWA context, suffers from two major problems. First, it
prioritizes a HWA \emph{only when} it is close to its deadlines, thus causing
the HWA to potentially miss deadlines. Second, it is \emph{not aware} of the
memory access characteristics of the different applications executing on
different agents (CPUs or HWAs), thus resulting in \emph{both} HWA deadline misses and low
CPU performance.

\textbf{Our goal}, in this work, is to design a memory scheduler that 1) meets
HWAs' deadlines and 2) at the same time maximizes CPU performance. To do so, we
design a scheduler that takes into account the differences in memory access
characteristics and demands of \emph{both} different CPU cores and HWAs.  Our
\emph{Simple QUality of service Aware memory Scheduler for Heterogeneous systems
(SQUASH)} is based on three key ideas.

First, to tackle the problem of HWAs missing their deadlines, SQUASH
prioritizes a HWA \emph{any time when it is not on track to meet its deadline}
(called \emph{Distributed Priority}),
instead of prioritizing it \emph{only} when close to a deadline . Effectively, our mechanism \emph{distributes} the priority of the
HWA over its deadline period, instead of clumping it towards the end of
the period.  This allows each HWA to receive \emph{consistent memory bandwidth}
throughout its run time.
%, unlike a scheme that prioritizes it only for a short amount of
%time close to a deadline.
Second, SQUASH exploits the heterogeneous memory access characteristics of
different \emph{CPU} applications and prioritizes HWAs over \emph{memory-intensive}
CPU applications \emph{even when} HWAs are on track to meet their deadlines.
The reason is that memory-intensive CPU applications' performance is not greatly
sensitive to additional memory latency.
%as also observed by ~\cite{atlas,tcm}.  
Hence, prioritizing HWAs over memory-intensive CPU applications enables
faster progress for HWAs and reduces the amount of time they are \emph{not} on track
to meet their deadlines. This, in turn, reduces the amount of time HWAs are
prioritized over \emph{memory-non-intensive} CPU applications that are
latency-sensitive, thereby achieving high overall CPU performance. Third,
SQUASH exploits the heterogeneous access characteristics of different
\emph{HWAs}.  We observe that a HWA with a \emph{short deadline period} needs a
short burst of high priority \emph{long enough} to ensure its few requests are served,
rather than consistent memory bandwidth. SQUASH achieves this by prioritizing
such HWAs for a short time period based on their estimated worst-case memory
access latency.

This paper makes the following main contributions.

\begin{itemize}

%\item{We present the characteristics of various HWAs and identify the problem
%of state-of-the-art memory schedulers not being able to both satisfy HWAs' QoS
%requirements and provide high CPU performance.}
\item{We identify a new problem: state-of-the-art memory schedulers
cannot both satisfy HWAs' QoS requirements and provide high CPU performance.}

%\item{We propose SQUASH, a new QoS-aware memory request scheduler that always
\item{We propose SQUASH, a new QoS-aware memory scheduler that \emph{always}
meets HWAs' deadlines while greatly improving CPU performance over the best previous scheduler.}
\item{We compare SQUASH to four different memory schedulers across a wide
variety of system configurations and workloads. We show that SQUASH
%improves average CPU performance by 14.8\% compared to the best-previous scheduler, while
improves CPU performance by 10.1\% compared to the best-previous scheduler, while
%always meeting all HWAs' deadlines.}
always meeting deadlines of all HWAs including GPUs.}

%improves average CPU performance by 13.9\% compared to the best-previous scheduler, while
%always meeting all HWAs' deadlines, thereby improving average HWA frame
%%rates by 10.9\%.}
%rates by 8.8\%. Compared to the HWA-friendly memory
%%rates by 8.8\%. Compared to the HWA-friendly 
%scheduler that always meets HWAs' deadlines, SQUASH improves
%CPU performance by 26.7\%.}
\end{itemize}

\section{Background}

In this section, we will first provide an overview of
heterogeneous SoC architectures and hardware accelerators (HWAs)
that are significant components in heterogeneous SoCs. Next, we
will provide a brief background on the organization of the DRAM
main memory and then describe the closest previous works on
main memory management and interference mitigation in
heterogeneous SoCs.
% that integrate CPUs and HWAs.

\subsection{Heterogeneous SoC Architecture} 
\begin{figure}
  \centering
  \includegraphics[scale=0.40, angle=0]{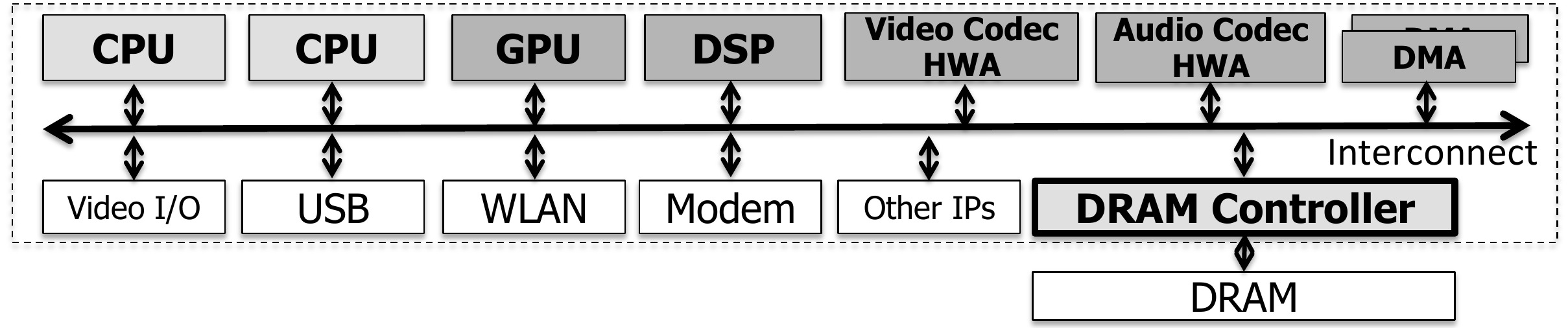}
%  \vspace{-2mm}
  \caption{Example heterogeneous SoC architecture}
  \label{fig:hetero-SoC}
\end{figure}

Modern SoCs are heterogeneous
architectures that integrate various kinds of processors.
Figure~\ref{fig:hetero-SoC} is an example of a typical high-end SoC designed for
smart phones~\cite{snapdragon,exnos}. The CPU is used to perform general purpose
computation. 
%\ifSQUEEZE
%\else
\TMP{There are multiple kinds of peripheral units such as video-I/O,
USB, WLAN controller, and modem.}
%\fi
HWAs are employed to accelerate various
functions. For instance, the GPU and Digital Signal Processor (DSP) are
optimized for graphics and digital signal processing respectively. Other
hard-wired HWAs are employed to perform video and audio coding at low power
consumption.
%The GPU and Digital Signal Processor (DSP) are fully programmable processing
%units optimized for graphics and digital signal processing. Besides such
%processing units, there are multiple kinds of peripheral units such as
%video-I/O, USB, WLAN controller, and modem. HWAs are employed to accelerate
%various functions. For instance, hard-wired HWAs are employed to perform video
%and audio coding at low power consumption.
Image recognition is another common function for which HWAs are
used~\cite{visconti2,mobileeye} because image recognition requires a large
amount of computation that embedded CPUs cannot perform in a timely
fashion~\cite{visconti2}. In such a heterogeneous SoC, the DRAM main memory is a
critical bottleneck between the CPU cores, HWAs, and DMA engines. Satisfying the
memory bandwidth requirements of all these requestors (or, agents) becomes a
major challenge. In this work, we will focus on managing memory bandwidth
between the CPU cores and HWAs with the goal of meeting deadline requirements
for HWAs while improving CPU performance.

\subsection{Hardware Accelerator Characteristics}
\label{sec:hwa_char}

%HWAs are designed for accelerating specific functions, thus having
%very different memory access patterns than CPU cores.
%There are a
%wide variety of HWAs integrated onto current SoCs.
%The functions that they accelerate are diverse and the implementations are also
%various. We will describe a 3x3 horizontal Sobel filter
%accelerator used for image recognition as an example.

\begin{figure}[h!]
  \centering
  \includegraphics[scale=0.40, angle=0]{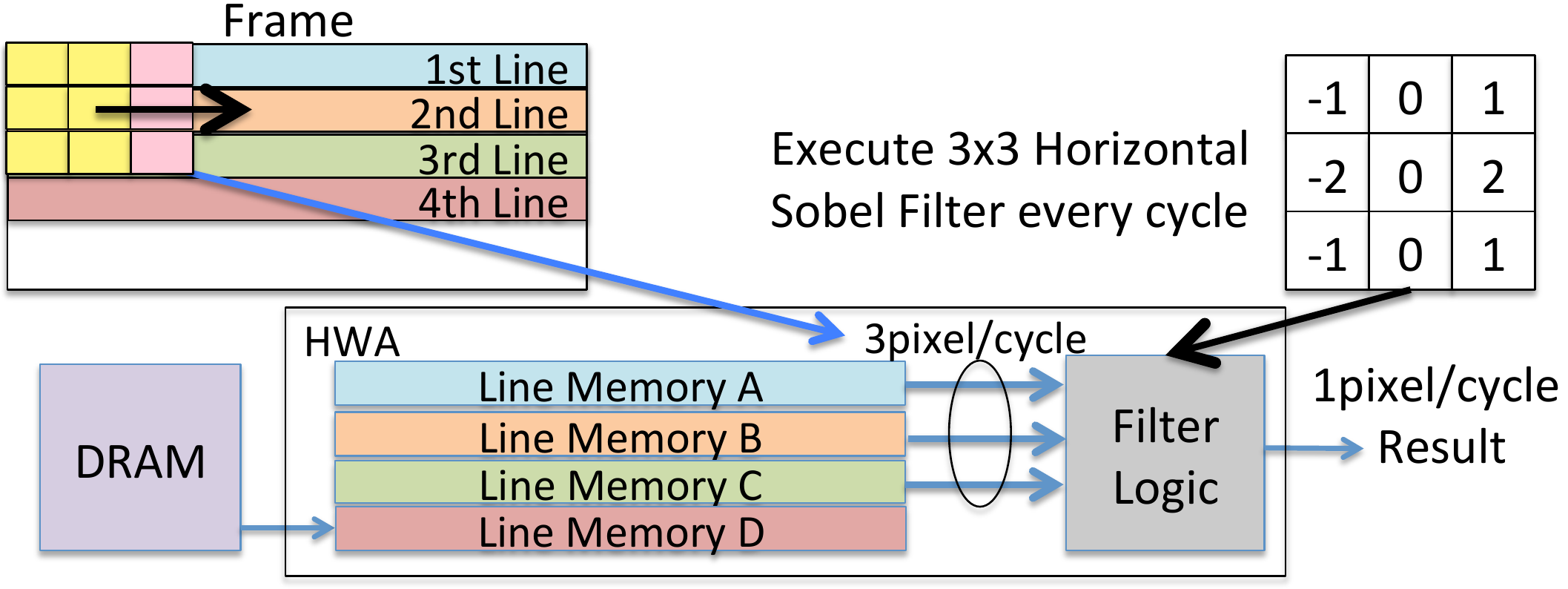}
  \caption{Typical implementation of a Sobel filter HWA}
  \label{fig:HWAexample}
\end{figure}

Modern SoCs consist of a wide variety of HWAs as each one is designed to
accelerate a specific function. 
\ifSQUEEZE
The functions that they accelerate are diverse
and the implementations also vary among different HWAs. 
As an example, we will first describe a typical implementation of a 
\emph{3x3 horizontal Sobel filter} accelerator~\cite{Sobel} (shown in Figure~\ref{fig:HWAexample}), 
which computes the gradient of an image for image recognition. 
%A Sobel filter~\cite{Sobel} computes the gradient of an image that is commonly used for edge detection. 
\else
The functions that they accelerate are diverse
and the implementations also vary among different HWAs. 
We will first describe a
\emph{3x3 horizontal Sobel filter} accelerator~\cite{Sobel} used for image recognition as an
example of HWAs which our work targets. 
A Sobel filter~\cite{Sobel} computes the gradient of an image that is commonly used for edge detection. 
Figure~\ref{fig:HWAexample}
shows a typical implementation of a Sobel filter accelerator that operates on
each 3x3 pixel block in a image.
\fi

%\kevin{Can you briefly describe what a Sobel filter is first? And what the 3x3
%window does (e.g., perform certain operations on 3x3 pixels in an image). }
The accelerator executes the Sobel filter on a target VGA
(640x480) image, at a target frame rate of 30 frames per second
(fps). 
A typical implementation for the filter uses \emph{line memory}
to take advantage of data access locality and hide the memory
access latency, as shown in Figure~\ref{fig:HWAexample}. The line
memory (consisting of lines A, B, C and D) can hold four lines,
each of size 640 pixels, of the target image. The filter operates
on three lines, at any point in time, while the next line is being
prefetched. For instance, the filter operates on lines A, B, and
C, while line D is being prefetched. After the filter finishes
processing lines A, B, and C, it operates on lines B, C, and D,
while line A is being prefetched. As long as the next line is
prefetched while the filter is operating on the three previous
lines, memory access latency does not affect performance. To meet
%a target performance (e.g., 30 fps), the filtering operation on a
a performance target (30 fps), the filtering operation on a
set of lines and the fetching of the next line have to be finished within
%69.44 $\micro$sec ($=1\;sec/30\;frames/480\;lines$). In this case,
%the \emph{period} of the HWA is 69.44 $\micro$sec and 
69.44 $\mu$sec ($=1\;sec/30\;frames/480\;lines$). In this case,
the \emph{period} of the HWA is 69.44 $\mu$sec and 
\ifSQUEEZE
the next line needs to be prefetched by the
end of the period (the \emph{deadline}).
\else
the requests
generated to prefetch the next line need to be completed by the
end of the period (the \emph{deadline}). 
\fi
Missing this deadline
causes the filtering logic to stall \ADD{and drop the frame being
processed}. As a result, it prevents the system from achieving the
performance target.

On the other hand, if the next-line prefetch is finished earlier
than the deadline, the prefetch of the line after that cannot be
initiated because the line memory can hold only one extra
prefetched line. Prefetching more than one line can overwrite the
contents of the line that is currently being processed by the
filter logic. 
%\ifSQUEEZE
%\ADD{In order to provision for more capacity to hold prefetched
%data, double buffers (e.g., frame buffers used in
%GPUs) are implemented in some HWAs.}
%\else
\TMP{In order to provision for more capacity to hold prefetched
data and hide memory latency better, double buffers (e.g., frame buffers used in
GPUs) are implemented in some HWAs.}
%\fi

\ifSQUEEZE
There are several HWAs with similar architectures employing line/frame
buffers, which are widely used in the media processing domain. 
HWAs for resizing an image~\cite{resizing} or feature extraction~\cite{HWA_sift,HWA_face}
use line buffers. HWAs for acoustic feature extraction~\cite{HWA_acoustic} use frame buffers. 
In all these HWAs, computing engines can only access line/frame
buffers and data is prefetched into these buffers from main
memory.
\else
There are several HWAs with similar architectures employing line buffers, that
are widely used in the media processing domain. HWAs for image processing and
recognition access images with a sliding window operation. The sliding window
operation scans a target image using a fixed size rectangular window. For
instance, HWAs have been built for resizing an image~\cite{resizing} using
bicubic interpolation that refers to the nearest 4x4 pixels window. HWAs for
feature extraction~\cite{HWA_sift} build up a Gaussian pyramid with a sliding
window of 25x25 pixels. Face detection can be performed using
HWAs~\cite{HWA_face} that extract feature data from the pixels in a window and
determine whether or not a face is included in the window. These HWAs are
implemented with line memories. Some HWAs for acoustic feature
extraction~\cite{HWA_acoustic} have frame buffers. In all these HWAs, computing
engines can only access line/frame buffers and HWAs fill data into buffers by
accessing DRAM.
\fi

Regardless of the types of buffers being used, one common
attribute across all these HWAs is that the amount of available
buffer capacity determines the \emph{deadline}, or \emph{period}, and how much
%data needs to be sent for each period of the HWA. For instance, in
data needs to be sent for each \emph{period}. For instance, in
the Sobel filter HWA example described above, the HWA requires
continuous bandwidth during each period (640 bytes for every 69.44
%$\micro$sec). As long as this continuous bandwidth is allocated,
$\mu$sec). As long as this continuous bandwidth is allocated,
the HWA is tolerant of memory latency. On the other hand, finishing the
HWA's memory accesses earlier than the deadline is wasteful,
%\ifSQUEEZE
%especially in a system with other agents such as CPUs, where
%the memory bandwidth can be better utilized to achieve higher
%performance of the other agents.
%\else
\TMP{especially in a system with other agents such as CPU cores, where
the memory bandwidth can be better utilized to achieve higher
overall performance for the other agents.}
%\fi

%\ADD{In the media processing areas where HWAs are widely applied,
%many kinds of HWAs are implemented with similar architecture.
%For example, HWAs for image processing and recognition access images with sliding window operation.
%The sliding window opperation scans target image with fixed sized rectangle
%windows (e.g., 24x24 pixels). The HWA~\cite{resizing} that resizes an image with
%bicubic interpolation computes an interpolated pixel with the nearest 16 (4x4) pixels.
%%The HWA can be implemented with 4 line memories and additional line memories for prefetching.
%The HWA for face detection~\cite{HWA_face} extracts feature data from pixels in the window and
%judges whether a face is included or not with the feature data. The HWA for feature extraction\cite{HWA_sift}
%builds up Gaussian pyramid with the sliding window of 25x25 pixels.
%In these cases, implementation with line memories whose capacity is
%greater than the vertical size of the window can hide the memory access latency.
%Another example is the HWA for acoustic feature extraction~\cite{HWA_acoustic} that has a frame buffer,
%which stores a fixed-sized audio data. Double frame buffer can hide the memory access latency even if
%the HWA shares DRAM with CPUs.}

As a result, a major challenge in today's heterogeneous SoCs is to ensure that
the HWAs get a consistent share of main memory bandwidth such that their
deadlines are met, while allocating enough bandwidth to the CPU cores to
achieve high CPU performance. This challenge is not solved by today's memory
schedulers which focus on either the HWAs or the CPUs.
As we will show in our evaluation (Section~\ref{sec:evaluation}), the HWA-friendly memory
scheduler that achieves almost 100\% deadline-met ratio for HWAs has 12\% lower
performance compared to the CPU-friendly scheduler that attains the highest CPU
performance without always meeting the deadlines. \textbf{The goal of our work} is \emph{to
both meet the HWAs' deadlines and maximize CPU performance}.

\subsection{DRAM Main Memory Organization}

% I modified this section to be top-down, starting with channels.
% Also, I made the description more high level because I think
% that suits a memory scheduling paper better. We can discuss if
% you have concerns or questions.

A typical DRAM main memory system is organized as a hierarchy of channels,
ranks, and banks. Each channel has its own address and data bus that operate
independently. A channel consists of one or more ranks. A rank, in turn,
consists of multiple banks. Each bank can operate independently and in parallel
with the other banks. However, all the banks on a channel share the address
and data buses of the channel.

Each bank consists of a 2D array (rows and columns) of cells.
When a piece of data is accessed from a bank, the
%entire row (~4KB) containing the piece of data is brought into a
entire row containing the piece of data is brought into a bank-internal
structure called the \emph{row buffer}. Any subsequent access to other data in
the same row can be served from the row buffer itself without incurring the
additional latency of accessing the array. Such an access is called a \emph{row
hit}. On the other hand, if the subsequent access is to data in a different
row, the array needs to be accessed and the new row needs to be brought into the
row buffer. Such an access is called a \emph{row miss}. A \emph{row miss} incurs
more than 2x the latency of a row hit~\cite{parbs,frfcfs}.

\subsection{Memory Management in Heterogeneous Systems}

%A large body of previous work has tackled the problem of managing
%memory bandwidth between applications in CPU-only multicore
%systems. The predominant direction a lot of these works took is
%memory request scheduling~\cite{frfcfs,stfm,parbs,fqm,atlas,tcm,mise}, reordering
%requests at the memory controller with the goal of mitigating
%interference between applications and improving system
%performance, fairness and Quality of Service (QoS). These
%techniques are designed for CPU-only systems and do not take into
%account the memory access characteristics and needs of other
%requestors such as HWAs. Some other previous
%works~\cite{predator,memguard} have proposed techniques to
%guarantee a minimum memory bandwidth and bound the maximum
%response time. However, these works do not take into account the
%%memory access demands of different requestors.
%memory access characteristics of different requestors.

Many previous works have tackled the problem of managing memory
bandwidth between applications in CPU-only multicore
systems~\cite{frfcfs,stfm,parbs,fqm,atlas,tcm,mise}. However, few previous works
have tackled the problem of memory management in heterogeneous systems, taking
%into account the memory access characteristics of the different requestors. One
into account the memory access characteristics of the different agents. One
previous work~\cite{armwhite} attempts to satisfy both high system performance
and QoS by acknowledging the differences in memory access characteristics
between CPU cores and other agents. They observe that CPU cores are latency
%\ifSQUEEZE
%sensitive, whereas the GPU is bandwidth
%sensitive with high memory latency tolerance. Therefore, they propose to
%prioritize CPU requests over GPU requests.
%\else
\TMP{sensitive, whereas the GPU and Video Processing Unit (VPU) are bandwidth
sensitive with high memory latency tolerance. Therefore, they propose to
prioritize CPU requests over GPU requests, while attempting to provide sufficient bandwidth to the GPU. }
%\fi
However, with such a scheme, the GPU cannot
always achieve its performance target when the CPU demands high
bandwidth~\cite{schedulingCPUGPU}. %Although always prioritizing GPU requests
%over CPU requests can meet the GPU's target, it degrades CPU performance.

%This work observed that CPU cores are
%latency sensitive, while the GPU and Video Processing Unit (VPU) that
%accelerates video decoding and encoding, are bandwidth sensitive and have better
%memory latency tolerance. Therefore, as long as sufficient bandwidth is
%consistently provided to the GPU and VPU over a period of time, they can meet
%their target performance. Based on observing these characteristics, they
%proposed techniques to regulate the outstanding GPU bandwidth demand and
%prioritize CPU requests over GPU requests. However, if CPU applications require
%large amounts of bandwidth, the GPU's target performance is often not
%satisfied~\cite{schedulingCPUGPU}. In contrast, always prioritizing the GPU over
%CPU cores can enable the GPU to meet its target frame rate. However, such a
%scheme would cause significant interference to the CPU cores.

In order to address this challenge of managing memory bandwidth
between the CPU cores and the GPU, a state-of-the-art technique
~\cite{schedulingCPUGPU} proposed to dynamically adjust memory
access priorities between the CPU cores and the GPU. This policy
compares current progress in terms of tiles rendered in a
frame (Equation~\ref{eq:FrameProgress}) against expected progress
in terms of time elapsed in a period
(Equation~\ref{eq:ExpectedProgress}) and adjusts priorities.

\begin{small}
  \begin{equation}
    CurrentProgress = \frac{Number\ of\ tiles\ rendered}{Number\ of\ tiles\ in\ the\ frame}
    \label{eq:FrameProgress}
  \end{equation}
  \begin{equation}
    ExpectedProgress = \frac{Time\ elapsed\ in\ the\ current\ frame}{Period\ for\ each\ frame}
    \label{eq:ExpectedProgress}
  \end{equation}
\end{small}

When {\it CurrentProgress} is greater than {\it ExpectedProgress}, the GPU is on
track to meet its target frame rate. Hence, GPU requests are given lower priority than
CPU requests. On the other hand, if {\it CurrentProgress} is less than or equal to
{\it ExpectedProgress}, the GPU is not on track to meet its target frame rate.
%\ifSQUEEZE
%In order to enable the GPU to make better progress, its requests
%are given the {\it same} priority as CPU requests. Only when the {\it
%ExpectedProgress} is greater than an {\it EmergentThreshold}
%(=0.90), is the GPU given higher priority than the CPU. 
%\else
\TMP{In order to enable the GPU to make better progress, its requests are given the
{\it same} priority as the CPU cores' requests. Only when the {\it
ExpectedProgress} is greater than an {\it EmergentThreshold} (=0.90), are the GPU's
requests given higher priority than the CPU cores' requests.}
%\fi
Such a policy
aims to preserve CPU performance, while still giving the GPU highest priority
%\ifSQUEEZE
%close to the end of a frame, 
%\else
\TMP{close to the time when a frame needs to be completed,}
%\fi
thereby providing better
QoS to the GPU than static prioritization policies. However, this policy, when
used within the context of a CPU-HWA system, is not adaptive enough, as we will
show next, to the diverse characteristics of different CPU applications and
HWAs.

\section{Motivation and Key Ideas}
\label{sec:motivation}

In this work, we examine heterogeneous systems that consist of multiple
HWAs and CPU cores executing applications with very diverse characteristics
(e.g., memory intensity and deadline requirements).
\MODIFIED{Although we have a different usage scenario from the previous work
that targets CPU-GPU systems by Jeong et al.~\cite{schedulingCPUGPU} (discussed
in the previous section), we adapt their scheduling policy in ~\cite{schedulingCPUGPU} and
apply it to manage memory bandwidth between CPU cores and HWAs since it
is the best previous work we know of in memory scheduling for heterogeneous systems.
We call this policy \emph{Dyn-Prio}.}

% but to understand the effectiveness of prior work on such a diverse system, we adapt a
%state-of-the-art technique, the dynamic priority policy proposed by Jeong et
%al.~\cite{schedulingCPUGPU} (discussed in the previous section) and apply it to
%manage memory bandwidth between CPU cores and HWAs. We call such a policy
%\emph{Dyn-Prio} for short.

Similar to GPUs' frame rate requirements, HWAs need to meet
deadlines. We target HWAs having soft deadlines, such as HWAs for
image processing and image recognition. A deadline miss for such
HWAs causes frames to be dropped. We re-define
\emph{CurrentProgress} and \emph{ExpectedProgress} in
Equations~\ref{eq:hwa_cur_prog} and \ref{eq:hwa_exp_prog},
respectively to capture HWAs' deadline requirements.

%Furthermore, like
%GPUs, HWAs also have higher memory latency tolerance, because of the presence of
%scratch pad memory such as line buffers and double buffers.
%
%A similar policy, as the one proposed by Jeong et
%al., in~\cite{schedulingCPUGPU} for CPU-GPU systems can be
%applied to manage memory bandwidth between CPU cores and HWAs,
%since HWAs have several characteristics that are similar to GPUs.
%Similar to GPUs' frame rate requirements, HWAs need to meet
%deadlines. Furthermore, like GPUs, HWAs also have higher memory
%latency tolerance, because of the presence of scratch pad memory
%such as line buffers and double buffers.

\begin{small}
  \begin{equation}
    CurrentProgress =\frac{\#\ of\ completed\ memory\ requests / period}{\#\ of\ total\ memory\ requests / period}
    \label{eq:hwa_cur_prog}
  \end{equation}
  \begin{equation}
    ExpectedProgress = \frac{Time\ elapsed\ in\ current\ period}{Total\ length\ of\ current\ period}
    \label{eq:hwa_exp_prog}
  \end{equation}
\end{small}

{\it CurrentProgress} for HWAs is defined as the fraction of the total
number of memory requests that have been completed. {\it ExpectedProgress} for
HWAs is defined in terms of the fraction of time elapsed during an
execution period.  \MODIFIED{In order to compute {\it CurrentProgress}, the
number of requests served during each period is needed. For several kinds of
HWAs, it is possible to precisely know this number due to two reasons. First,
as described in Section~\ref{sec:hwa_char}, a lot of HWAs for media processing
access media data in a streaming manner, resulting in a
predictable/prefetch-friendly access stream. Second, when a HWA is implemented
with a line-/double-buffered scratchpad, all the data required for the next
set of computations need to be prefetched into the scratchpad to meet a target
performance because the compute engines can only access the scratchpad. In
%this scenario, the number of memory requests in a deadline period can be
this scenario, the number of memory requests in a period can be
estimated in a fairly straightforward manner based on the amount of data that
are required to be prefetched.}
%Therefore, we think this definition can apply a lot of kinds of HWAs.}

%For fixed-function HWAs, it is possible to precisely know
%the number of memory requests that need to be served during each
%period, because most HWAs prefetch a fixed amount of data for the
%next set of computations.

%HWAs completely isolate the series of memory accessses from computation
%with scratch-pad memory, at the beginning of the series of memory accesses, the number of
%memory requests are fixed.

%\begin{small}
%\begin{eqnarray*}
%CurrentProgress = \frac{\#\ of\ finished\ memory\ requests / period}{\#\ of\ total\ memory\ requests / period}\\
%\label{eq:hwa_cur_prog}
%%\end{equation}
%%\begin{equation}
%ExpectedProgress = \frac{Time\ elapsed\ in\ the\ current\ period}{Total\ length\ of\ current\ period}
%\label{eq:hwa_exp_prog}
%\end{eqnarray*}
%\end{small}
We observe that there are two major problems with \mbox{Dyn-Prio}
when used in a CPU-HWA context.
First, it only prioritizes a HWA over CPU cores close to the HWA's
deadline (i.e., after 90\% \emph{ExpectedProgress} has been made).
Prioritizing HWAs only when the deadline is very close can cause
deadline misses because the available memory bandwidth in the
remaining time before the deadline may not be able to
sustain the required memory request rates of all the HWAs and CPUs.
We will explain this problem in greater detail in Section~\ref{sec:distprio}.
\ifSQUEEZE
Second, Dyn-Prio is designed for a simple heterogeneous CPU-GPU system
\else
Second, Dyn-Prio is designed for a simple heterogeneous system
with only one GPU and some CPU cores 
\fi
and is not designed to
%consider the access characteristics of different applications accessing
%memory in a heterogeneous system executing different kinds of
consider the access characteristics of different applications
in a heterogeneous system executing different kinds of
applications on different kinds of agents, either on the CPUs or on the HWAs.
As we will explain in
Section~\ref{sec:obs1} and \ref{sec:obs2}, application-unawareness
misses opportunities to improve system performance because
different applications executing on CPUs and HWAs have different
latency tolerance and bandwidth requirements.

\subsection{Key Idea 1: Distributed Priority}
\label{sec:distprio}

To address the first problem where HWAs sometimes miss their deadlines, we
propose a simple modification.  A HWA enters a state of urgency and is given the
highest priority anytime when its \emph{CurrentProgress} is less than or equal
to \emph{ExpectedProgress}. We call such a scheme \emph{Distributed Priority}
(\emph{Dist-Prio} for short). Using \emph{Dist-Prio} distributes a HWA's
priority over its deadline period, rather than clumping it close to a deadline.
This allows HWAs to receive consistent memory bandwidth and make \emph{steady progress}
throughout their run time.

\begin{figure}[t]
  \centering
  \subfloat[CPU-A (Memory-non-intensive application)]{\includegraphics[scale=0.45, angle=0]{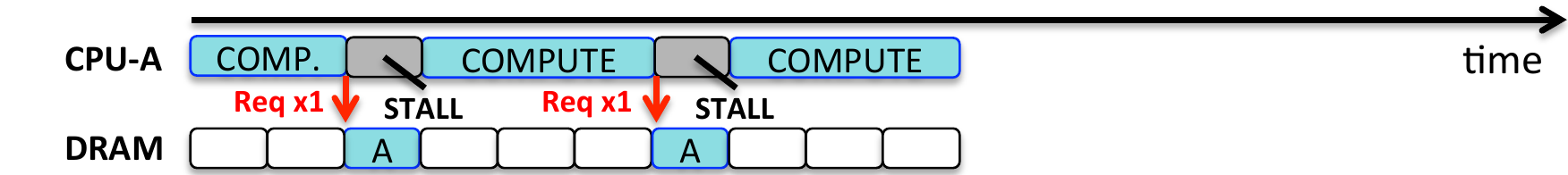}\label{fig:singlemaster_cpua}}
  \vspace{-2mm}
  \\
  \subfloat[CPU-B (Memory-intensive application)]{\includegraphics[scale=0.45, angle=0]{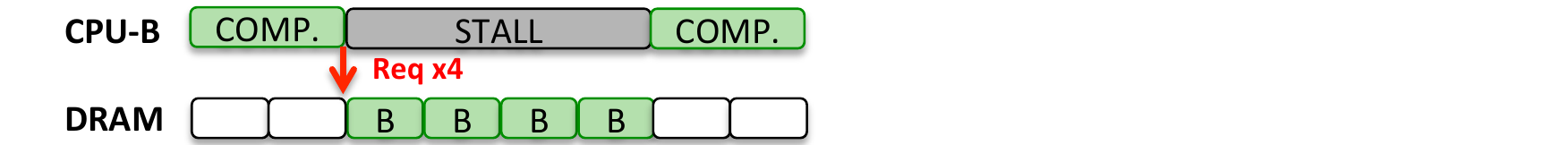}\label{fig:singlemaster_cpub}}
  \vspace{-2mm}
  \\
  \subfloat[HWA]{\includegraphics[scale=0.45, angle=0]{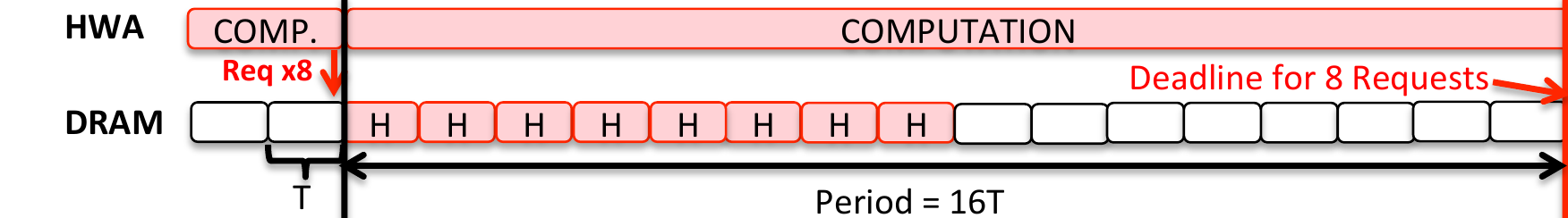}\label{fig:singlemaster_hwa}}
  \caption{Memory service timeline example when each agent is executed alone}
  \label{fig:singlemaster}
\end{figure}

To illustrate the benefit of such a policy, we study an example system
with two CPU cores and a HWA. Figure~\ref{fig:singlemaster} shows the execution
%timelines along with the number of generated requests when each agent (HWA or
timelines when each agent (HWA or
CPU core) executes alone. In this example, \ca has low memory intensity and
generates very few memory requests. In contrast, \cb has high memory intensity
and generates more memory requests than \ca does. \hwa has double buffers and
generates 8 prefetch requests for the data to be processed in the next period.
For ease of understanding, we assume all these requests are destined to the
same bank and each memory request takes T cycles (no distinction between row
%hits and misses). The length of the \hwa's execution period is 16T. When a
hits and misses). The length of the \hwa's period is 16T. When a
\emph{Dyn-Prio} scheme with an \emph{EmergentThreshold} of 0.9 is employed to
schedule these requests, the \hwa is given highest priority only for the last
two time units, starting at time 14T. Until then, the CPU cores' requests are
treated on par with the HWA's requests. Such a short amount of time is not
sufficient to finish serving the HWA's requests in this example. 
\ifSQUEEZE
\else
Our evaluations
across real HWAs show that such a policy causes deadline misses for many HWAs. On average, only
50\% of the deadlines are met for a particular HWA (i.e., MAT-HWA) across the
evaluated 80 workloads.\footnote{Our evaluation methodology and system
configuration are described in Section~\ref{sec:methodology}} This results in
a frame rate of only 15 fps, as opposed to the required frame rate of 30 fps.
\fi

%This might not be enough to finish serving all the HWA's requests.
%Hence, employing the \emph{Dyn-Prio} scheme can potentially cause
%deadline misses.
%\ADD{In our evaluations across 80
%workloads~\footnote{Our evaluation methodology and system
%configuration are described in Section~\ref{sec:methodology}},
%only half of the deadlines are met for the MAT-HWA when a
%\emph{Dyn-Prio} is employed, where the HWA's requests are
%prioritized only during the last 10\% of a deadline period. This
%results in a frame rate of only 16 fps, as against the required
%frame rate of 30 fps.}

\showcaptionsetup{subfloat}
\captionsetup[subfloat]{captionskip=-0.5pt}
\captionsetup[subfloat]{nearskip=-0.5pt}
%\captionsetup[figure]{skip=-1pt}

%\begin{figure}[t]
%  \centering
%  \begin{minipage}{0.45\textwidth}
%  \subfloat[CPU-A (Memory-non-intensive application)]{\includegraphics[scale=0.42, angle=0]{figs/fig_timing_alone_cpua}\label{fig:singlemaster_cpua}}
%  \end{minipage}
%  \begin{minipage}{0.45\textwidth}
%  \subfloat[CPU-B (Memory-intensive application)]{\includegraphics[scale=0.42, angle=0]{figs/fig_timing_alone_cpub}\label{fig:singlemaster_cpub}}
%  \end{minipage}
%  \begin{minipage}{0.45\textwidth}
%  \subfloat[HWA]{\includegraphics[scale=0.42, angle=0]{figs/fig_timing_alone_hwa}\label{fig:singlemaster_hwa}}
%  \end{minipage}
%  \vspace{-2mm}
%  \caption{Memory service timeline example when each agent is executed alone}
%  \label{fig:singlemaster}
%\end{figure}

\begin{figure}[t]
  \centering
%  \begin{minipage}{0.45\textwidth}
  \subfloat[Distributed priority]{\includegraphics[scale=0.42, angle=0]{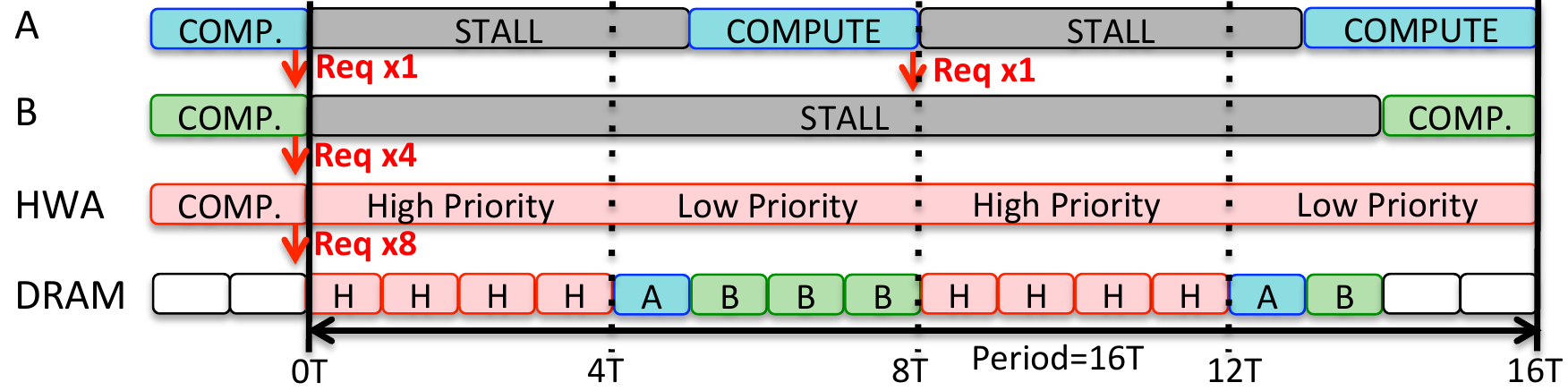}\label{fig:schedulingexample-dynamic}}
%  \end{minipage}
%  \begin{minipage}{0.45\textwidth}
  \\
  \subfloat[Application-aware distributed priority]{\includegraphics[scale=0.42, angle=0]{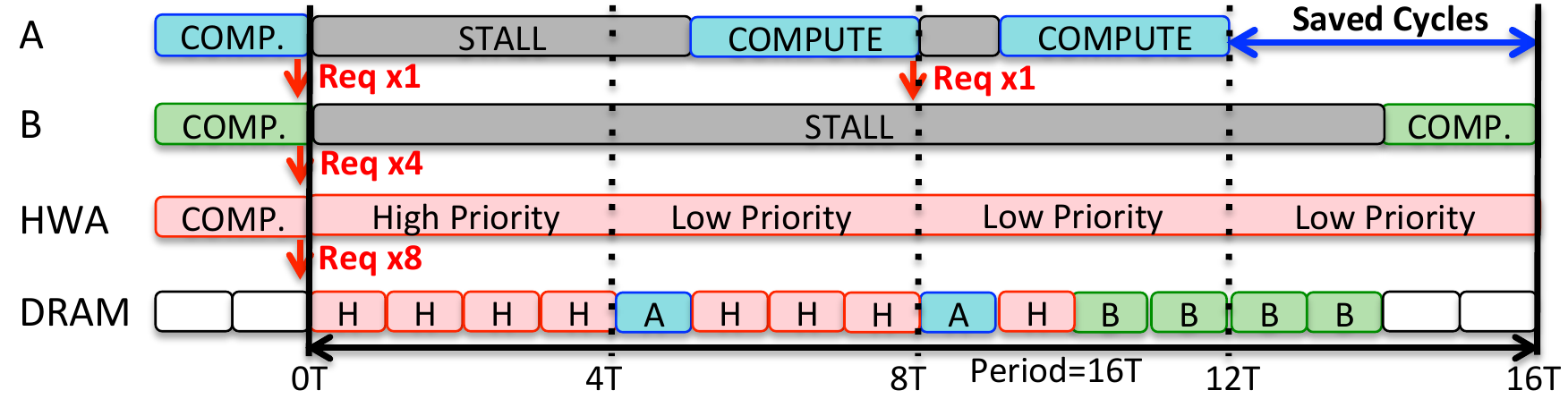}\label{fig:schedulingexample-squash}}
%  \end{minipage}
  \vspace{-2mm}
  \caption{Memory service timeline example when all agents execute together}
  \label{fig:schedulingexample}
\end{figure}

Figure~\ref{fig:schedulingexample-dynamic} illustrates the scheduling order of
requests from a \emph{single system} with a HWA (\hwa) and two CPU cores (\ca
and \cb) using our proposed \emph{Dist-Prio} scheme. It prioritizes the HWA \emph{any time}
when it is not on track to meet its deadline. Among the CPU cores, the low
memory-intensity \ca is prioritized over the high memory-intensity \cb. At the
beginning of the deadline period, since both {\it CurrentProgress} and {\it
ExpectedProgress} are zero and equal, \hwa is deemed as urgent and is given
higher priority than the CPU cores. Hence, during the first 4T cycles, only
\hwa's requests are served. After 4T cycles, {\it CurrentProgress} is 0.5 and
{\it ExpectedProgress} is 0.25.  Hence, \hwa is deemed as not urgent and is
given lower priority than the CPU cores. Requests from both CPU cores are
served from cycles 4T to 8T. After 8T cycles, since both {\it CurrentProgress}
and {\it ExpectedProgress} are 0.50, \hwa is deemed as urgent again and its
remaining requests are completed.
Hence, Dist-Prio enables the HWA to meet its deadlines while also achieving
high CPU performance.

\subsection{Key Idea 2: Application-Aware Scheduling for CPUs}
\label{sec:obs1}

We observe that when HWAs are given higher priority than CPU cores, they interfere
with all CPU cores' memory requests. For instance, in
Figure~\ref{fig:schedulingexample-dynamic}, during cycles 8T to
12T, \hwa stalls both \ca and \cb. Furthermore, higher the
memory intensity of the HWAs, more the memory bandwidth they need
to make sufficient progress to meet their deadlines, exacerbating
the interference. We propose to tackle this shortcoming based on
the observation that memory-intensive applications do not
experience significant performance degradation when HWAs are
prioritized over them.
%first observation.
%
%\textbf{Observation 1:} {\it Memory-intensive applications do not
%experience significant performance degradation when HWAs are
%prioritized over them.}

%Previous works~\cite{atlas,tcm} also have observed that applications with low
Applications with low memory-intensity are more sensitive to memory latency,
since these applications generate few memory requests, and quick service of
these few requests enables such applications to make good forward
progress.\footnote{This was also observed by some previous works and utilized in
the context of multi-core memory scheduling.~\cite{atlas,tcm}} On the other
hand, applications with high-memory-intensity often have a large number of
outstanding memory requests and spend a significant fraction of their execution
time stalling on memory. Therefore, delaying high-memory-intensity
applications' requests does not impact their performance significantly. Based on
this observation, we propose to prioritize HWAs' memory requests over those of
high-memory-intensity applications even when HWAs are making sufficient progress
and are \emph{not} in a state of urgency. Such a prioritization scheme reduces
the number of cycles when HWAs are deemed \emph{urgent} and prioritized over
\emph{memory-non-intensive} CPU applications that are latency-sensitive, thereby
improving their performance.

Figure~\ref{fig:schedulingexample-squash} illustrates the benefits of such an
application-aware distributed priority scheme for the same set of requests shown
in Figure~\ref{fig:schedulingexample-dynamic}. The request schedule remains the
same during the first 4T cycles. At time 4T, the {\it CurrentProgress} is 0.5
and {\it ExpectedProgress} is 0.25. Since {\it CurrentProgress} is greater than
{\it ExpectedProgress}, \hwa is not deemed urgent and \ca's request is
prioritized over \hwa's requests. However, \hwa is \emph{still} prioritized over
\cb that has high memory-intensity, enabling faster progress for \hwa. As a
result, at time 8T, {\it CurrentProgress} is 0.875, which is greater than {\it
ExpectedProgress}. As such, the HWA is \emph{still} deemed \emph{not urgent},
unlike in the distributed priority case (in
Figure~\ref{fig:schedulingexample-dynamic}). Hence, the latency-sensitive \ca's
requests are served earlier. Thus, this key idea enables faster progress for
\ca, as can be seen from Figure~\ref{fig:schedulingexample-squash}, and results
in higher CPU performance.

\subsection{Key Idea 3: Application-Aware Scheduling for HWAs}
\label{sec:obs2}
%\ifSQUEEZE
%Monitoring a HWA's progress and prioritizing it when it is not on track to meet
%its deadline is an effective mechanism to ensure consistent bandwidth to HWAs
%that have fairly long periods. However, such a scheme is
%not effective for HWAs with short periods since it is
%difficult to ensure that these HWAs are deemed urgent and receive enough
%priority for sufficiently long periods of time within a short deadline period.
%\else
\TMP{Monitoring a HWA's progress and prioritizing it when it is not on track to meet
its deadline is an effective mechanism to ensure consistent bandwidth to HWAs
that have fairly long periods (infrequent deadlines). However, such a scheme is
not effective for HWAs with short periods (frequent deadlines) since it is
difficult to ensure that these HWAs are deemed urgent and receive enough
priority for sufficiently long periods of time within a short deadline period.}
%\fi
Specifically, a short deadline provides little time for all requests and
causes the HWAs to be more susceptible to interference from other HWAs and CPU
cores. We evaluated a heterogeneous system with two accelerators (HWA-A and
HWA-B) that have vastly different period lengths (i.e., 63041 and 5447 cycles)
and bandwidth requirements (i.e., 8.32GB/s and 475MB/s) employing our previously
two key ideas. Our results show that HWA-A meets all its deadlines whereas
HWA-B, on average, misses a deadline every 2000 execution periods.

To help enable better deadline-met ratios for HWAs with short deadlines, we make
the following two observations that lead to our third key idea. First, HWAs with
short deadline periods can be enabled to meet their deadlines by giving them a
short burst of highest priority very close to the deadline. Second, prioritizing
short-deadline-period HWAs does not cause much interference to other requestors
%because these HWAs generate few memory requests, consuming a small amount of
because these HWAs consume a small amount of
bandwidth. Based on these observations, we propose to estimate the
$\mathit{WorstCaseLatency}$ for a memory access and give a short-deadline-period HWA
highest priority for $\mathit{WorstCaseLatency*NumberOfRequests}$ cycles close to its
deadline.

\section{Mechanism}

In this section, we describe the details of SQUASH, our proposed memory
scheduling mechanism to manage memory bandwidth between CPU cores and HWAs,
using the three key ideas described in Section~\ref{sec:motivation}.
First, we describe a scheduling policy to prioritize between HWAs with
long deadline periods and CPU applications, with the goal of enabling the
long-deadline-period HWAs to meet their deadlines while improving CPU
performance (Section~\ref{sec:longdeadline}). Second, we describe how SQUASH
enables HWAs with short deadline periods to meet their deadlines
(Section~\ref{sec:shortdeadline}). Third, we present a combined scheduling policy for
long and short-deadline-period HWAs (Section~\ref{sec:overall-scheduling}). Finally, we
 describe a modification to our original scheduling policy to
probabilistically change priorities between long-deadline-period HWAs and CPU
applications to enable higher fairness for memory-intensive CPU applications
(Section~\ref{sec:prob-switching}), which results in the final SQUASH mechanism.

\textbf{Overview:} SQUASH categorizes HWAs as long and short-deadline-period
statically based on their deadline period. \kevin{Based on what metric? And
why?}A different scheduling policy is employed for each of these two
categories, since they have different kinds of bandwidth demand. For the
long-deadline-period accelerators (\emph{LDP-HWAs} for short), SQUASH monitors
their progress periodically and appropriately prioritizes them, enabling them
to get sufficient and consistent bandwidth to meet their deadlines (Section~\ref{sec:distprio}). For the
short-deadline-period accelerators (\emph{SDP-HWAs} for short), SQUASH gives
them a short burst of highest priority close to each deadline, based on
worst-case access latency calculations (Section~\ref{sec:obs2}). SQUASH
also treats memory-intensive and memory-non-intensive CPU applications
differently with respect to their priority over HWAs (Section~\ref{sec:obs1}).

\subsection{Long-Deadline-Period HWAs vs. CPU Applications}
\label{sec:longdeadline}

To make scheduling decisions between LDP-HWAs and CPU applications, SQUASH
employs the \emph{Distributed Priority (Dist-Prio)} scheme as previously described
in Section~\ref{sec:distprio}, monitoring each LDP-HWA's progress every
\emph{SchedulingUnit}. SQUASH prioritizes LDP-HWAs over CPUs only when LDP-HWAs
become \emph{urgent} under either of the following conditions: 1)
$\emph{CurrentProgress} \leq \emph{ExpectedProgress}$ or 2)
$\emph{ExpectedProgress} > \emph{EmergentThreshold}$.

%SQUASH monitors each HWA's {\it CurrentProgress} and {\it ExpectedProgress}, as
%defined in Equations~\ref{eq:hwa_cur_prog} and~\ref{eq:hwa_exp_prog} respectively, every
%{\it SchedulingUnit}. At the end of a {\it SchedulingUnit}, {\it
%CurrentProgress} and {\it ExpectedProgress} are compared.  When {\it
%CurrentProgress} is less than or equal to {\it ExpectedProgress}, the HWA is
%classified as urgent. Otherwise, it is classified as non-urgent. Furthermore,
%apart from the progress based urgency computation, after an {\it
%EmergentThreshold} (90\% of the deadline period in our experiments), a
%long-deadline-period HWA is classified as urgent.
%
%
%CPU applications' memory-intensity is monitored during each {\it
%SchedulingUnit}.  At the end of a {\it SchedulingUnit},
%applications are classified as memory-non-intensive or
%memory-intensive based on their calculated memory-intensity using
%the classification mechanism employed in TCM~\cite{tcm}.

CPU applications' memory-intensity is monitored and
applications are classified as memory-non-intensive or
memory-intensive periodically based on their calculated memory-intensity using
the classification mechanism borrowed from ~\cite{tcm}.
Note that other mechanisms can also be employed to perform this
classification.\footnote{Also note that even though we borrow the classification
mechanism of~\cite{tcm} to categorize memory-intensive and memory-non-intensive
applications, the problem we solve and the scheduling policy we devise are very
different from those of~\cite{tcm}}

Based on this classification, SQUASH schedules requests
at the memory controller in the following priority order (lower number
indicates higher priority):
%(lower number indicates higher priority):

\begin{enumerate}
\item{Urgent HWAs}
\item{Memory-non-intensive CPU applications}
\item{Non-urgent HWAs}
\item{Memory-intensive CPU applications}
\end{enumerate}

Based on our first observation in Section~\ref{sec:obs1},
HWAs' requests are prioritized over memory-intensive CPU
applications' requests, even when the HWAs are not deemed urgent
since memory-intensive applications are not latency-sensitive.

\subsection{Providing QoS to HWAs with Short Deadline Periods}
\label{sec:shortdeadline}

Although using Dist-Prio can provide consistent bandwidth to LDP-HWAs
to meet their deadlines, SDP-HWAs do not get enough memory bandwidth to meet
their deadlines (as we described in our third key idea in
Section~\ref{sec:obs2}). In order to enable SDP-HWAs to meet their deadlines, we
propose to give them a short burst of high priority close to a deadline using
estimated worst-case memory latency calculations.

\kevin{Links back to my previous comment on using Dist-Prio for SDP? Isn't this
scheme essentially changing the emergency condition based on the worst-case mem
latency?}

%While HWAs with long-deadline-periods can often get a consistent
%share of memory bandwidth and meet their deadlines, through a
%scheme that monitors progress and appropriately assigns
%priorities, HWAs with short deadline periods do not get a
%consistent share of memory bandwidth with such a scheme (as we
%described in our second observation in Section~\ref{sec:obs2}).
%In order to enable HWAs with short deadline periods to meet their
%deadlines, we propose to give them a short burst of high priority
%close to a deadline, based on worst case latency calculations.

\noindent\textbf{Estimating worst case access latency.} In the worst case, all
requests from a SDP-HWA would access different rows in the same bank. In this
case, all such requests are serialized and each request takes tRC - the minimum
time between two DRAM row ACTIVATE operations.\footnote{Accesses to different
rows within the same bank have to be spaced apart by a fixed timing of tRC based
on the DRAM specification~\cite{jedec-ddr3}.} Therefore, in order to serve the
requests of an SDP-HWA before its deadline, it needs to be deemed urgent and
given the highest priority over all other requestors for $tRC*NumberOfRequests$
for each period, which we call it \emph{Urgent Period Length (UPL)}.

For example, when a HWA outputs 16 requests every 2000 ns period and tRC is 50
ns, the HWA is deemed urgent and given the highest priority for 800 ns +
$\alpha$ during each period, where $\alpha$ is the waiting time for the
in-flight requests to finish. Furthermore, finishing a HWA's requests much
earlier than the deadline is wasteful, since doing so does not improve the HWA's
performance any further. Hence, this highest priority period can be at the end
of the HWA's deadline period. For instance, in the previous example, the HWA is
given highest priority ($2000-(800+\alpha)$) ns after a deadline period starts.

\noindent\textbf{Handling multiple short-deadline-period HWAs.}
The scheme discussed above does not consider the scenarios when there are
multiple \sdphwa, which could overlap with each other during the high priority
cycles, resulting in deadline misses.
We propose to address this concern using the following
mechanism:

\begin{enumerate}

\item
SQUASH calculates the urgent period length (UPL) of each SDP-HWA $x$ as: \\
$\mathit{UPL}(x) = tRC*TheNumberOfRequests(x)$

\item
Among the urgent short-deadline-period HWAs, the HWAs with shorter
deadline periods are given higher priority.

\item SQUASH extends the urgent period length of each SDP-HWA $x$ further by
taking into account all the SDP-HWAs that have higher priority (i.e., shorter
deadline period) than $x$. This ensures that each HWA is allocated enough
cycles for its urgent period. When calculating how long we should extend a
\mbox{SDP-HWA $x$'s} $\mathit{UPL}$, we calculate how many deadline periods ($N_i$)
of each higher priority \mbox{SDP-HWA} ($i$) can overlap with the $\mathit{UPL}$
of $x$: $N_i = \lceil(\mathit{UPL}(x)/Period(i))\rceil$. We then calculate the
total length of high-priority $\mathit{UPL}$, $\mathit{HP\mathchar`-UPL}(i)$,
resulting from $N_i$ high-priority deadline periods:
$\mathit{HP\mathchar`-UPL}(i) = N_i * \mathit{UPL}(i)$, which we use to add to
the current SDP-HWA's UPL. In summary, the final extension function for each
SDP-HWA $x$ is: $\mathit{UPL}(x) = \Sigma_i (\mathit{HP\mathchar`-UPL}(i)) +
\mathit{UPL}(x)$, for all HWAs $i$ that have higher priority than $x$.

%\item
%SQUASH calculates the urgent period length of each SDP-HWA $x$ as: \\
%$WC\mathchar`-MAL(x) = tRC*TheNumberOfRequests(x)$
%
%\item
%Among the urgent short-deadline-period HWAs, the HWAs with shorter
%deadline periods are given higher priority.
%
%\item SQUASH extends the urgent period length of each SDP-HWA $x$ further by
%taking into account which other SDP-HWAs have higher priority than $x$. This
%ensures that each HWA is allocated enough cycles during when it enters its urgent
%period. The extension function for each $x$ is shown below: \\
%$WC\mathchar`-MAL(x) = \Sigma_i (\lceil(WC\mathchar`-MAL(x)/Period(i))\rceil *
%WC\mathchar`-MAL(i)) + WC\mathchar`-MAL(x)$, for all HWAs $i$ that have
%higher priority than $x$. This extension provisions for the fact that the
%$WC\mathchar`-MAL(x)$ of a HWA x can overlap with more than one
%$WC\mathchar`-MAL(i)$ of another HWA i.

\end{enumerate}

\subsection{Overall Scheduling Policy}
\label{sec:overall-scheduling} Combining the mechanisms described in
Sections~\ref{sec:longdeadline} and~\ref{sec:shortdeadline}, SQUASH schedules
requests in the following order (lower number indicates higher priority) and
priority level within each group is also provided in parentheses:

\begin{enumerate}
\item
\textbf{Urgent HWAs in the short deadline period group} (Higher
priority to shorter deadline HWAs)
\item
\textbf{Urgent HWAs in the long deadline period group} (Higher
priority to earlier deadline HWAs)
\item
\textbf{Non-memory-intensive CPU applications} (Higher priority to
lower memory-intensity applications)
\item
\textbf{Non-urgent HWAs in the long deadline period group} (Higher
priority to earlier deadline HWAs)
\item
\textbf{Memory-intensive CPU applications} (Application priorities are
shuffled as in ~\cite{tcm})
\item
\textbf{Non-urgent HWAs in the short and long deadline period group}
(Higher priority to earlier deadline HWAs)
\end{enumerate}

\ifOPTION
The current scheduling order allows HWAs to receive high priority when they
becomes urgent (i.e., not meeting their expected progress). This prevents them 
from missing deadlines due to interference from CPU applications. Memory-intensive CPU applications
($Group5$) are always ranked lower than memory-non-intensive CPU applications
($Group3$) and LDP-HWAs ($Group2,4$).
\else
In this order, memory-intensive CPU applications
($Group5$) are always ranked lower than non-memory-intensive CPU applications
($Group3$) and LDP-HWAs ($Group2,4$). 
\fi
This can potentially always deprioritize memory-intensive
applications when the memory bandwidth is only enough to serve memory-non-intensive
applications and HWAs. To ensure memory-intensive applications receive sufficient memory
bandwidth to make progress, we employ a clustering mechanism that only allocates
a fraction of total memory bandwidth (called \emph{ClusterFactor}) to the memory-non-intensive 
group~\cite{tcm}.

%\ifSQUEEZE
%As we explain in Section~\ref{sec:distprio}, the initial state of all HWAs is
%urgent. When HWAs meet their expected progress, they make the transition to the
%non-urgent state. 
%\else
\TMP{As explained in Section~\ref{sec:distprio}, the initial state of all HWAs is
urgent when using \emph{Dist-Prio}. When HWAs meet their expected progress, they make the
transition to the non-urgent state, allocating more bandwidth to CPU
applications or other HWAs. }
%\fi
Non-urgent SDP-HWAs are always in $Group6$. Non-urgent LDP-HWAs, however, can be in
either $Group4$ or $Group6$. They are assigned to $Group6$ only
when they first transition to the non-urgent state, but are assigned to $Group4$
when they re-enter the non-urgent state later on. The rationale is that LDP-HWAs
do not need to be prioritized over memory-intensive CPU applications ($Group5$)
if they are already receiving memory bandwidth such that they continuously meet
their expected progress, without ever transitioning back to the urgent state
again, throughout the period.
\ifOPTION
{This kind of a priority scheme enables LDP-HWAs to make progress while not over-consuming memory
bandwidth and enables memory-intensive CPU applications to achieve
higher performance.}
\else
Using $Group6$ enables the memory-intensive CPU applications to achieve
higher performance.
\fi

\subsection{Probabilistic Switching of LDP-HWAs' Priorities}
\label{sec:prob-switching}

By using the scheduling order described in the previous section, we observe that
memory-intensive applications experience unfair slowdowns due to interference
from non-urgent \ldphwas in some workloads. To solve this problem, we propose a
mechanism to probabilistically prioritize memory-intensive applications over
non-urgent LDP-HWAs, switching priorities between $Group4$ and $Group5$. Each
LDP-HWA $x$ has a probability value $Pb(x)$ that is controlled based on its
request progress every epoch ({\it SwitchingUnit}).
Algorithm~\ref{alg:sched_pb} shows how requests are scheduled based on $Pb(x)$.
With a probability of $Pb(x)$, memory-intensive applications are prioritized
over LDP-HWA $x$ to enable higher fairness. Algorithm~\ref{alg:cntl_pb} shows
the periodic adjustment of $Pb(x)$ using empirically determined steps. We use a
larger decrement step than the increment step because we want to quickly reduce
the priority of memory-intensive applications in order to increase the HWA's
bandwidth allocation when it is not making sufficient progress. This
probabilistic switching helps ensure that the memory-intensive CPU applications
are treated fairly.
% as our evaluations in Section~\ref{sec:evaluation} show.

\vspace{-2mm}
\begin{algorithm}[t!]
\caption{Scheduling using $Pb(x)$}
\label{alg:sched_pb}
\small
\begin{algorithmic}
\STATE \textbf{With a probability $Pb(x)$}: \\
Memory-intensive applications $>$ Long-deadline-period HWA $x$
\STATE \textbf{With a probability $(1-Pb(x))$}: \\
Memory-intensive applications $<$ Long-deadline-period HWA $x$
\end{algorithmic}
\end{algorithm}
\vspace{-2mm}

\begin{algorithm}[t!]
\caption{Controlling $Pb(x)$ for LDP-HWAs}
\label{alg:cntl_pb}
\small
\begin{algorithmic}
\STATE \textbf{Initialization}: $Pb(x)$ = 0\\
%\STATE \textbf{Adjusting}: (every {\it SwitchingUnit})\\
\textbf{\underline{Every {\it SwitchingUnit:}}}
\IF{ {\it CurrentProgress} > {\it ExpectedProgress}}
%\STATE $Pb(x) = Pb(x) + Pb_{inc}$ ($Pb_{inc} = 1\%$ in our experiments)
\STATE $Pb(x) \mathrel{{+}{=}} Pb_{inc}$ ($Pb_{inc} = 1\%$ in our experiments)
\ELSIF{ {\it CurrentProgress} < {\it ExpectedProgress}}
%\ELSE
%\IF{ {\it CurrentProgress} < {\it ExpectedProgress}}
%\STATE $Pb(x) = Pb(x) - Pb_{dec}$ ($Pb_{dec} = 5\%$ in our experiments)
\STATE $Pb(x) \mathrel{{-}{=}} Pb_{dec}$ ($Pb_{dec} = 5\%$ in our experiments)
\ELSE
\STATE $Pb(x) = Pb(x)$
\ENDIF
\end{algorithmic}
\end{algorithm}
\vspace{-5mm}

\ifSQUEEZE
\else
\subsection{Estimating Memory Requirements of HWAs}

SQUASH is based on the assumption that the number of memory requests served
during a deadline period for a HWA can be known ahead of time. As explained in
Section~\ref{sec:motivation}, we can precisely determine this number for many
kinds of HWAs that use scratchpad
memory~\cite{resizing,HWA_sift,HWA_face,HWA_acoustic,mra}, especially in the
media processing space, which is a major segment of today's SoC market.
\fi

\section{Implementation and Hardware Cost}

SQUASH requires hardware support to monitor HWAs' current and expected progress
and schedule memory requests accordingly. To track current progress, the memory
controller counts the number of completed requests during a deadline period. If
there are multiple memory controllers, they send their recorded counter values
to a centralized meta-controller every \emph{SchedulingUnit}, similar to
~\cite{atlas,tcm}. If HWAs access shared caches, the number of completed
requests at the shared caches is sent to the meta-controller.
Table~\ref{tab:storage} lists the major counters required for the
meta-controller over a baseline TCM scheduler~\cite{tcm}, the state-of-the-art
application-aware scheduler for multi-core systems, which we later
provide comparison to. The request counters are used to track current progress,
whereas the cycle counters are used to compute expected progress. $Pb$ is the
probability that determines priorities between long-deadline-period HWAs and
memory-intensive applications. A 4-byte counter is sufficient to denote each of
these quantities. Hence, the total counter overhead is 20 bytes for a
long-deadline-period HWA and 12 bytes for a short-deadline-period HWA.

\begin{table}[h!]
\vspace{-3mm}
\footnotesize
  \centering
  \begin{tabular}{|l|l|}
    \hline
    \multicolumn{2}{|l|}{For long-deadline-period HWAs} \\
    \hline
    \textbf{Name} & \textbf{Function} \\
    \hline
    {\it Curr-Req} & Number of requests completed in a
    deadline period\\
    \hline
    {\it Total-Req} & Total number of requests completed in a
    deadline period\\
    \hline
    {\it Curr-Cyc} & Number of cycles elapsed in a deadline period\\
    \hline
    {\it Total-Cyc} & Total number of cycles in a deadline period\\
    \hline
    {\it Pb} & Probability when memory-intensive applications \\
    & $>$ long-deadline-period HWAs \\
    \hline
    \hline
    \multicolumn{2}{|l|}{For short-deadline-period HWAs} \\
    \hline
    \textbf{Name} & \textbf{Function} \\
    \hline
    {\it Priority-Cyc} & Indicates when the priority is transitioned to high \\
    \hline
    {\it Curr-Cyc} & Number of cycles elapsed in a deadline period\\
    \hline
    {\it Total-Cyc} & Total number of cycles in a deadline period\\
    \hline
  \end{tabular}
\vspace{-2mm}
\caption{Storage required for SQUASH}
\label{tab:storage}
\vspace{-3mm}
\end{table}

%\vspace{-5mm}
{\it Total-Req} and {\it Total-Cyc} are set by the system software based on the
specifications of HWAs. If these parameters are fixed for the target HWA, the
software sets up these registers at the beginning of execution. If these
parameters vary for each period, the software
updates them at the beginning of each period. {\it Curr-Cyc} is incremented
every cycle. {\it Curr-Req} is incremented every time a request is completed (at
the respective memory controller). At the end of every {\it SchedulingUnit}, the
meta controller computes {\it ExpectedProgress} and {\it CurrentProgress} using
these accumulated counts, in order to determine how urgent each
long-deadline-period HWA is. For the short-deadline-period HWAs, their state of
urgency is determined based on {\it Priority-Cyc} and {\it Curr-Cyc}. {\it
Priority-Cyc} is set by the system software based on the HWAs' specifications.
This information is used along with $Pb$ to determine the scheduling order
across all HWAs and CPU applications. Once this priority order is determined,
the meta-controller broadcasts the priority to the memory controllers, and the
memory controllers schedule requests based on this priority order, similar to
other application-aware memory schedulers~\cite{stfm,parbs,atlas,tcm}.

\newcommand{\aloneipc}{\textrm{IPC}_{i}^{\scriptstyle{alone}}}
\newcommand{\sharedipc}{\textrm{IPC}_{i}^{\scriptstyle{shared}}}

\section{Methodology}
\label{sec:methodology}
\subsection{System Configuration}
We use an in-house cycle-level simulator to perform our evaluations. We model a
system with eight x86 CPU cores and four HWAs for our main evaluation. 
To avoid starving CPU cores or
HWAs, we allocate half of the memory request buffer entries that hold memory
requests to CPU cores and the other half to HWAs. Unless stated
otherwise, our system configuration is as shown in Table~\ref{sim_env}.

\begin{table}[h!]
\vspace{-3mm}
%\scriptsize
\footnotesize
  \centering
  \begin{tabular}{|l|l|}
    \hline
    CPU & 8 Cores, 2.66GHz, 3-wide issues\\
        &  128 entry instruction window, 16 MSHRs/core \\
    \hline
    L1Cache & Private, 2 way, 32 KB, 64 Byte Line \\
    \hline
    L2Cache & Shared, 16 way, 4 MB, 64 Byte Line \\
    \hline
    HWA & 4 HWAs \\
    \hline
    \multirow{2}{*}{\begin{minipage}{0.5in}DRAM\end{minipage}} & DDR3-1333
    (9-9-9) \cite{micron-ddr3}, 300 request buffer entries \\
    & 2 channels, 1 rank per channel, 8 banks per rank \\
    \hline
  \end{tabular}
  \vspace{-2mm}
  \caption{Configuration of the simulated system}
  \label{sim_env}
\end{table}
\vspace{-0.3in}

\subsection{Workloads for CPUs} 
We construct 80 multiprogrammed workloads from
the SPEC CPU2006 suite \cite{spec2006}, TPC \cite{tpc}, and the NAS parallel
benchmark suite \cite{nas}. We use Pin \cite{pin} with PinPoints \cite{pinpoint}
to extract representative phases. We classify CPU benchmarks into two
categories, memory-intensive and memory-non-intensive, based on the number of
last-level cache misses per thousand instructions (MPKI). If an application's
MPKI is greater than 5, it is classified as a memory-intensive application.
Otherwise, it is classified as memory-non-intensive. We then construct five
intensity categories of workloads based on the fraction of memory-intensive
benchmarks in a workload: 0\%, 25\%, 50\%, 75\%, and 100\%. Each category
consists of 16 workloads.

%We construct 80 multiprogrammed workloads for Conf-H \MODIFIED{and 30 multiprogrammed workloads for 
%Conf-G and Conf-HG} from the SPEC CPU2006 suite \cite{spec2006}, TPC \cite{tpc}, and the NAS parallel
%benchmark suite \cite{nas}. We use Pin \cite{pin} with PinPoints \cite{pinpoint}
%to extract representative phases. We simulate for
%200 million CPU cycles.
%We classify CPU benchmarks into two
%categories, memory-intensive and memory-non-intensive, based on the number of
%last-level cache misses per thousand instructions (MPKI). If an application's
%MPKI is greater than 5, it is classified as a memory-intensive application.
%Otherwise, it is classified as memory-non-intensive. We then construct five
%intensity categories of workloads based on the fraction of memory-intensive
%benchmarks in a workload: 0\%, 25\%, 50\%, 75\%, and 100\%. \MODIFIED{Each category
%consists of 16 workloads for Conf-H. For Conf-G and Conf-HG, we use three
%categories of 0\%, 50\%, 100\% and each category consists of 10 workloads.}

\subsection{Hardware Accelerators}\label{sec:HWAs}
We use five kinds of HWAs designed for image processing and recognition, for our
evaluations, as described in Table~\ref{tab:hwa_env}. The target frame rate for
the HWAs is 30 fps.  The image processing HWA (IMG-HWA) performs filter
processing on input RGB images of size 1920x1080. We assume that IMG-HWA
performs filter processing on one frame for 1/30 sec with double buffers.
Hessian HWA (HES-HWA) and Matching HWA (MAT-HWA) are designed for Augmented
Reality (AR) systems~\cite{mra}. HES-HWA accelerates the fast Hessian detector
that is executed in SURF (Speed up Robust Features) \cite{surf}, which is used
to detect interesting points and generate descriptors. MAT-HWA accelerates the
operation of matching descriptors generated by SURF against those in a database.
The implementation of HES-HWA and MAT-HWA are based on~\cite{mra}. Their
configuration parameters are as shown in Table~\ref{tab:hwa_env}. 
We evaluate HES-HWA and MAT-HWA for three different
configurations. The periods and bandwidth requirements of the HWAs are different depending on
the configuration.
%\ifSQUEEZE
%\else
\TMP{We assume that
the result of the MAT-HWA is output to a register in the HWA.}
%\fi
Resize HWA (RSZ-HWA) and Detect HWA (DET-HWA) are used for face detection. Their
implementations are based on a library that uses Haar-Like
features~\cite{haarlike}, included in Open-CV~\cite{opencv}. RSZ-HWA shrinks the
target frame recursively in order to detect differences in sizes of faces and
generates integral images. DET-HWA detects faces included in the resized image.
Because the target image is shrunk recursively over each frame, the HWAs'
periods are variable. 
%We evaluate HES-HWA and MAT-HWA for three different
%configurations. The periods and bandwidth requirements of the HWAs are different depending on
%the configuration. 
The HES-HWA and DET-HWA are categorized into
the short-deadline-period group and the others into the long-deadline-period
group.

Based on the implementations of the HWAs, we build trace-generators that
simulate memory requests from the HWAs. All HWAs have fixed access patterns
throughout the simulation run. We evaluate two mixes of HWAs, Config-A and
Config-B, with each CPU workload, as shown in Table~\ref{tab:hwa_env}. 
%\ifSQUEEZE
%\else
\TMP{Config-B
includes HWAs that dynamically change their bandwidth requirements and deadlines over time,
which we use to evaluate the adaptivity of different schedulers. }
%\fi
We simulate for
200 million CPU cycles.
The size of memory requests from HWAs is 64 bytes and the
number of outstanding requests from each HWA to the memory is at most 16.

\begin{table}[t!]
\vspace{-2mm}
\footnotesize
%\scriptsize
  \centering
\setlength{\tabcolsep}{.3em}
  \begin{tabular}{|l|l|l|l|}
    \hline
    & \textbf{Period} & \textbf{Bandwidth} & \textbf{Scratchpad} \\
    \hline
    IMG-HWA & 33 ms & 360 MB/s & double buffer (1 frame x 4) \\
    \hline
    HES-HWA(32) & 2 us & 478 MB/s & line buffer (32 lines)\\
    \cline{1-3}
    HES-HWA(64) & 4 us & 329 MB/s & 30 lines for computation\\
    \cline{1-3}
    HES-HWA(128) & 8 us & 224 MB/s & 2 lines for prefetch\\
    \hline
    MAT-HWA(30) & 23.6 us & 8.32 GB/s & double buffer (4 KB x 4)\\
    \cline{1-3}
    MAT-HWA(20) & 35.4 us & 5.55 GB/s & 4 KB x 2 for query\\
    \cline{1-3}
    MAT-HWA(10) & 47.2 us & 2.77 GB/s & 4 KB x 2 for database\\
    \hline
    RSZ-HWA & 46.5 us - & 2.07 GB/s - & double buffer (1 frame x 4) \\
    & 5183 us & 3.33 GB/s & \\
    \hline
    DET-HWA & 0.8 us - & 1.60 GB/s - & line buffer (26 lines) \\
    & 9.6 us & 1.86 GB/s & 24 lines for computation \\
    \hline
    \hline
    & \multicolumn{3}{l|}{\textbf{Parameters}} \\
%    \hline
%    IMG-HWA & \\
    \hline
    HES-HWA(N) & \multicolumn{3}{l|}{ image size: 1920 x 1080, max filter size:
    30,} \\
    ~\cite{mra} & \multicolumn{3}{l|}{N entries operated at the same time}\\
    \hline
    MAT-HWA(M) & \multicolumn{3}{l|}{ 3000 interesting points (64 dimension) per
    image,} \\
    ~\cite{mra} & \multicolumn{3}{l|}{matching M images} \\
    \hline
    RSZ-HWA & \multicolumn{3}{l|}{ image size: 1920 x 1080, scale factor : 1.1,} \\
    DET-HWA~\cite{opencv} & \multicolumn{3}{l|}{ 24 x 24 window } \\
    \hline
    \hline
    & \multicolumn{3}{l|}{\textbf{Configuration}} \\
    \hline
    Config-A & \multicolumn{3}{l|}{ IMG-HWA x 2, MAT-HWA(30), HES-HWA(32) } \\
    \hline
    Config-B & \multicolumn{3}{l|}{ MAT-HWA(20), HES-HWA(32), RSZ-HWA, DET-HWA } \\
    \hline
  \end{tabular}
  \vspace{-2mm}
  \caption{Configuration of the HWAs}
  \label{tab:hwa_env}
\vspace{-2mm}
\end{table}

\subsection{System with a GPU}
In addition to our CPU-HWA evaluations, we also evaluate CPU-GPU
and CPU-GPU-HWA systems. The specification of the GPU we model is
800 MHz, 20 cores and 1600 operations/cycle, which is similar to
the AMD Radeon 5870 specification \cite{amd-radeon}. The GPU does
not share caches with CPUs. The CPU-GPU-HWA system has four memory
channels and four HWAs, whose configuration is the same as Config-A
in Section~\ref{sec:HWAs}. The other system parameters are the
same as the CPU-HWA system. We collect GPU traces from GPU
benchmarks and games, as shown in Table~\ref{tab:gpu-bench}, with
a proprietary GPU simulator. The target frame rate of all GPU
benchmarks is 30 fps. We set the GPU benchmarks' deadline to 33.3
msec (= 1 frame). 
%If the GPU finishes all memory requests before
%a deadline, the GPU becomes idle until the deadline.  After the
%deadline, the GPU resumes and iterates the trace. 
We measure the number of memory requests included in each trace in advance and
use this number to calculate \emph{CurrentProgress}. We simulate 
30 CPU-GPU and CPU-GPU-HWA workloads respectively.

%Original longer version
%In addition to our CPU-HWA evaluations, we also evaluate CPU-GPU
%and CPU-GPU-HWA systems. The specification of the GPU we model is
%800 MHz, 20 cores and 1600 operations/cycle, which is similar to
%the AMD Radeon 5870 specification \cite{amd-radeon}. The GPU does
%not share caches with CPUs. The CPU-GPU-HWA system has four memory
%channels and four HWAs whose configuration is the same as Config-A
%in Section~\ref{sec:HWAs}. The other system parameters are the
%same as the CPU-HWA system. We collect GPU traces from GPU
%benchmarks and games, as shown in Table~\ref{tab:gpu-bench}, with
%a proprietary GPU simulator. Target frame rate of all GPU
%benchmarks is 30 fps. GPU memory accesses are collected after
%filtering GPU cache accesses. Each trace has memory requests
%included for one frame. We set GPU benchmarks' deadline to 33.3
%msec (= 1 frame). If the GPU finishes all memory requests before
%a deadline, the GPU becomes idle until the deadline.  After the
%deadline, the GPU resumes and iterates the trace. We measure the
%number of memory requests included in each trace in advance and
%use this number to calculate \emph{CurrentProgress}. We simulate 
%30 CPU-GPU and CPU-GPU-HWA workloads respectively.

%\vspace{-3mm}
\begin{table}[h!]
%\scriptsize
\footnotesize
  \centering
  \begin{tabular}{|l|l||l|l|}
    \hline
    Name & Description & Name & Description \\
    \hline
    \hline
    Bench & 3D mark & Game03 & Shooting Game 3 \\
    \hline
    Game01 & Shooting Game 1 & Game04 & Adventure Game \\
    \hline
    Game02 & Shooting Game 2 & Game05 & Role-playing Game \\
    \hline
  \end{tabular}
  \vspace{-2mm}
  \caption{GPU benchmarks}
  \label{tab:gpu-bench}
\end{table}

%\vspace{-7mm}

%\begin{table}[h!]
%\scriptsize
%  \centering
%  \begin{tabular}{|l|l|l|}
%    \hline
%    Name & Description & Target fps\\
%    \hline
%    \hline
%    Bench & 3D mark & 30 \\
%    \hline
%    Game01 & Shooting Game 1 & 30 \\
%    \hline
%    Game02 & Shooting Game 2 & 30 \\
%    \hline
%    Game03 & Shooting Game 3 & 30 \\
%    \hline
%    Game04 & Adventure Game & 30 \\
%    \hline
%    Game05 & Role-playing Game & 30 \\
%    \hline
%  \end{tabular}
%  \caption{GPU benchmarks}
%  \label{tab:gpu-bench}
%\end{table}

\subsection{Performance Metrics} We measure CPU performance with the
commonly-used \emph{Weighted Speedup (WS)}~\cite{weighted-speedup,ws} metric. We
measure fairness using the \emph{Maximum Slowdown} metric \cite{max-slowdown,
atlas, tcm}. For HWAs, we use two metrics: the \emph{DeadlineMetRatio} and
\emph{frame rate} in \emph{fps}, frames per second. We assume that if a deadline
is missed in a frame, the corresponding frame is dropped (and we calculate frame
rate accordingly).

%We use three metrics to evaluate CPU performance. The {\it
%WeightedSpeedup} metric \cite{weighted-speedup,ws} is used to
%measure system performance and the {\it MaximumSlowdown} metric
%\cite{max-slowdown, atlas, tcm} is used to measure unfairness. The
%{\it HarmonicSpeedup} metric \cite{harmonic-speedup} is used a
%measure of balance between system performance and fairness.
%
%\begin{small}
%  \begin{eqnarray*}
%    {\it Weighted Speedup} & = & \Sigma_i \frac{\sharedipc}{\aloneipc}\\
%    {\it Harmonic Speedup} & = & N/\left(\Sigma_i\frac{\aloneipc}{\sharedipc}\right)\\
%    {\it Maximum Slowdown} & = & \textrm{max}\left(\frac{\aloneipc}{\sharedipc}\right)\\
%  \end{eqnarray*}
%\end{small}
%
%For HWAs, we use two metrics: the {\it DeadlineMeetRatio} and
%frame rate. We assume that if a deadline is missed in a frame,
%the missed frame is skipped.

\subsection{Parameters of the Evaluated Schedulers}
Unless otherwise stated, for SQUASH, we set the {\it SchedulingUnit} to 1000 CPU cycles
and {\it SwitchingUnit} to 500 CPU cycles. 
%The {\it EmergentThreshold} is 0.9 (90\%). 
For TCM~\cite{tcm}, we use a \emph{ClusterFactor} of 0.2
and a shuffling interval of 800 cycles and {\it QuantumLength} of
1M cycles.

\section{Evaluation}\label{sec:evaluation} We compare SQUASH with previously
proposed schedulers: 1) FRFCFS~\cite{frfcfs} and TCM with static priority
(FRFCFS-St and \mbox{TCM-St}) where the HWA always has higher priority than all
CPU cores and 2) FRFCFS with \emph{Dyn-Prio} (FRFCFS-Dyn), which employs the
dynamic priority mechanism~\cite{schedulingCPUGPU}. \MODIFIEDISCA{We evaluate
two variants of the FRFCFS-Dyn mechanism with different {\it EmergentThreshold}
values. First, we use an {\it EmergentThreshold} value of 0.9 for all HWAs
(FRFCFS-Dyn0.9), which is designed to achieve high CPU performance. Second, in
order to achieve high deadline-met ratios for the HWAs, we sweep the value of
the {\it EmergentThreshold} from 0 to 1.0 at the granularity of 0.1 (see
Section~\ref{sec:sweeping_th} for more details) and choose a different threshold 
value shown in Table~\ref{tab:emergent_threshold} for each HWA (FRFCFS-DynOpt)
such that a deadline-met ratio greater than 99.9\% and a frame rate greater than
27 fps (90\% of target frame rate) are achieved.} For SQUASH, we use an {\it
EmergentThreshold} value of 0.8 for all HWAs.
%We explain the evaluation of sweeping {\it EmergentThreshold} in
%Section~\ref{sec:sweeping_th}.

%\vspace{-1mm}
\begin{table}[h!]
%\vspace{-2mm}
\scriptsize
  \centering
  \begin{tabular}{|c|c|c|c|c|c|c|}
    \hline
    \multicolumn{3}{|c|}{Config-A} & \multicolumn{4}{c|}{Config-B} \\
    \hline
    HES & MAT & IMG & HES & MAT & DET & RSZ \\
    \hline
    0.2 & 0.2 & 0.9 & 0.5 & 0.4 & 0.5 & 0.7 \\
    \hline
  \end{tabular}
  \vspace{-2mm}
  \caption{EmergentThreshold for FRFCFS-Dyn}
  \label{tab:emergent_threshold}
  \vspace{-3mm}
\end{table}
%\vspace{-1mm}

%\vspace{-6mm}
%We use {\it EmergentThreshold} values of 0.9
%and 0.3 for FRFCFS-Dyn.

%Schedulers with dynamic priority adjusts priority dynamically based on the progress of the HWA \cite{schedulingCPUGPU}.
%In order to anaylyze benefits of SQUASH, we evaluate schedulings that employ only \emph{Dist-Prio}
%to FRFCFS (SQUASH-FDist) and TCM (SQUASH-TDist).
%FRFCFS with dynamic distributed priority,
%(FRFCFS-DynD) TCM with static priority (TCM-St), TCM with dynamic distributed priority (TCM-DynD).
% When {\it CurrentProgress} is greater than {\it ExpectedProgress}, FRFCFS-Dyn sets the priority of a HWA to the same as CPUs, but FRFCFS-DynD and TCM-DynD
%set the priority of HWA higher than CPUs.

\begin{figure}[ht!]
  \centering
  \includegraphics[scale=0.25, angle=270]{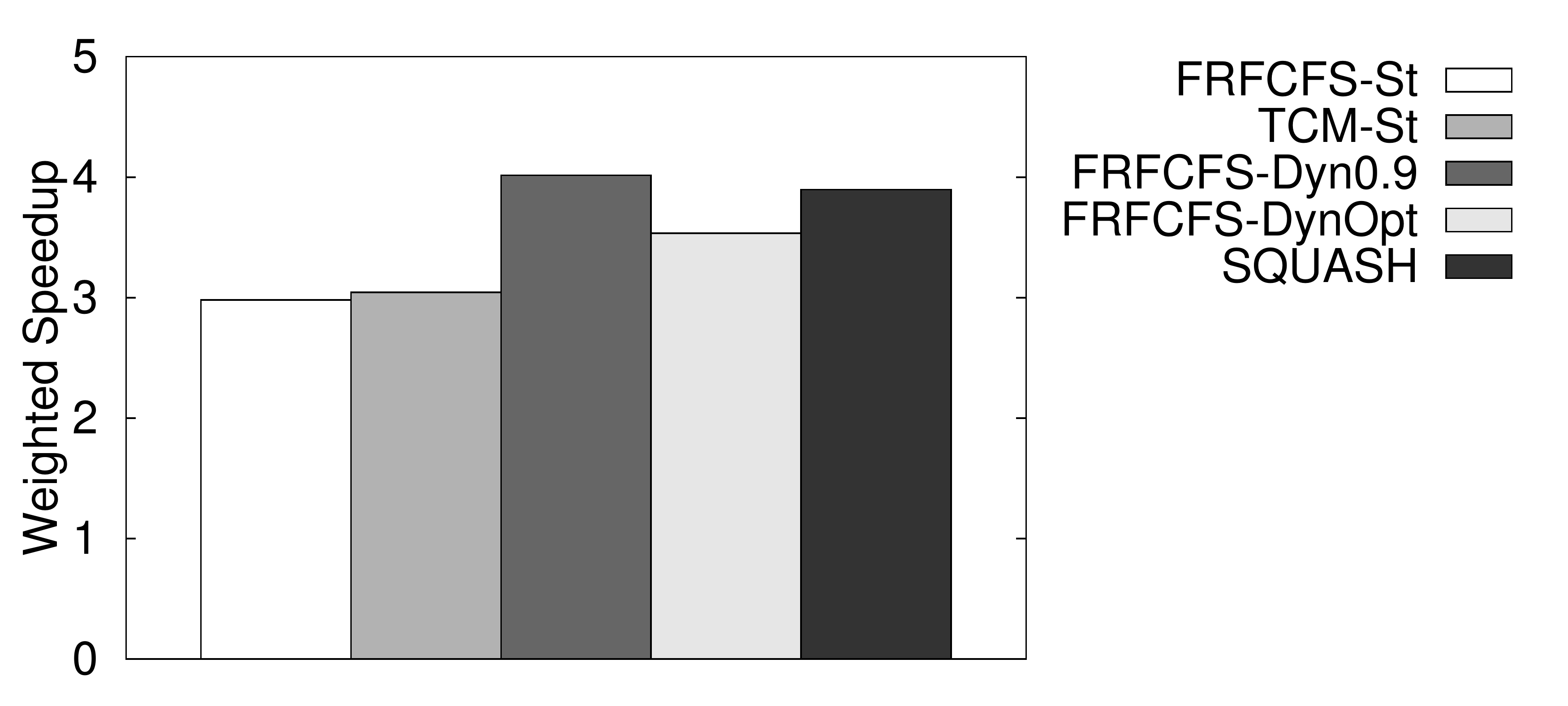}
%  \begin{minipage}{0.32\textwidth}
%    \flushleft
%    \includegraphics[scale=0.22, angle=270]{plots/main_avg_unnorm_hs}
%  \end{minipage}
%  \begin{minipage}{0.23\textwidth}
%    \flushleft
%    \includegraphics[scale=0.17, angle=270]{plots/main_avg_unnorm_ms}
%  \end{minipage}
%  \vspace{-3mm}
  \caption{System performance}
  \label{plot:average_performance}
\end{figure}

%\subsection{CPU-Fixed function HWA Results} 
Figure~\ref{plot:average_performance} shows the
average system performance across all 80 workloads, using both
Config-A and B. Table~\ref{tab:average_hwa_performance} shows the deadline-met
ratio and frame rate of four types of HWAs. We do not show IMG-HWA because it
has a 100\% deadline-met ratio with all schedulers.
%We calculate frame rate by
%assuming that a deadline miss causes the corresponding frame to be dropped.

%\vspace{-4mm}

%\vspace{-5mm}

%\begin{figure}[ht!]
%  \vspace{-7mm}
%  \flushleft
%  \begin{minipage}{0.23\textwidth}
%    \flushleft
%    \includegraphics[scale=0.17, angle=270]{plots/main_avg_unnorm_ws}
%  \end{minipage}
%%  \begin{minipage}{0.32\textwidth}
%%    \flushleft
%%    \includegraphics[scale=0.22, angle=270]{plots/main_avg_unnorm_hs}
%%  \end{minipage}
%  \begin{minipage}{0.23\textwidth}
%    \flushleft
%    \includegraphics[scale=0.17, angle=270]{plots/main_avg_unnorm_ms}
%  \end{minipage}
%  \vspace{-2mm}
%  \caption{System performance and fairness}
%  \label{plot:average_performance}
%\end{figure}
%\vspace{-4mm}

\begin{table}[h]
%\vspace{-5mm}
%\scriptsize
\footnotesize
\centering
\setlength{\tabcolsep}{.60em}
    \begin{tabular}{lcccc}
      \toprule
%      \multirow{2}{*}{\textbf{\begin{minipage}{0.5in}Scheduling Algorithms\end{minipage}}} &
      \multirow{2}{*}{\textbf{Scheduling Algorithms}} &
      \multicolumn{4}{c}{\textbf{Deadline-Met Ratio (\%) / Frame Rate (fps)}} \\
       & HES & MAT & RSZ & DET \\
      \midrule

    FRFCFS-St & 100 / 30 & 100 / 30 & 100 / 30 & 100 / 30 \\
    TCM-St & 100 / 30 & 100 / 30 & 100 / 30 & 100 / 30 \\
    FRFCFS-Dyn0.9 & 99.40 / 15.38 & 46.01 / 15.28 & 97.98 / 25.19 & 97.14 / 16.5 \\
    FRFCFS-DynOpt & 100 / 30 & 99.997 / 29.72 & 100 / 30 & 99.99 / 25.5 \\
%    SQUASH-FDist & 99.91 / 17.81 & 99.999 / 29.81 & 100 / 30 & 99.72 / 23.06 \\
%    SQUASH-TDist & 99.94 / 16.97 & 99.999 / 29.91 & 100 / 30 & 99.74 / 23.06 \\
%    SQUASH & 100 / 30 & 100 / 30 & 100 / 30 & 100 / 30 \\
    %SQUASH-Prob & 100 / 30 & 100 / 30 & 100 / 30 & 100 / 30 \\
    SQUASH & 100 / 30 & 100 / 30 & 100 / 30 & 100 / 30 \\

      \bottomrule
    \end{tabular}
\vspace{-2mm}
\caption{Deadline-met ratio and frame rate of HWAs}
\label{tab:average_hwa_performance}%
\end{table}

We draw three major observations. First, FRFCFS-St and TCM-St always prioritize
HWAs, achieving a 100\% deadline-met ratio. However, always prioritizing the
HWAs' requests results in low CPU performance. Second, the FRFCFS-Dyn policy either
achieves high CPU performance or high deadline-met ratio depending on the value
of the {\it EmergentThreshold}. When {\it EmergentThreshold} is 0.9, the HWAs
are not prioritized much, causing them to miss deadlines. However, CPU
performance is high. 
\MODIFIEDISCA{On the other hand, when we use optimized values of {\it
EmergentThreshold} (FRFCFS-DynOpt), the HWAs are prioritized enabling them to
meet almost all their deadlines, but at the cost of CPU performance. Third,
SQUASH achieves comparable performance to FRFCFS-Dyn-0.9 and 10.1\% better
system performance than FRFCFS-DynOpt, while achieving a deadline-met ratio of
100\%. 
%This comes at the cost of 44.5 \% higher unfairness, compared to
%FRFCFS-DynOpt, from prioritizing HWAs over memory-intensive applications. 
We conclude that SQUASH achieves both high CPU performance and 100\% QoS for
HWAs. In the next section, we present a breakdown of the benefits from the
different components of SQUASH.}

% Original Longer Version
%We draw three major observations. First, FRFCFS-St and TCM-St always prioritize
%HWAs, achieving a 100\% deadline-met ratio. However, always prioritizing the
%HWAs' requests results in low CPU performance, as shown in
%Figure~\ref{plot:average_performance}. Second, the FRFCFS-Dyn policy either
%achieves high CPU performance or high deadline-met ratio depending on the value
%of the {\it EmergentThreshold}. When {\it EmergentThreshold} is 0.9, the HWAs
%are not prioritized much, causing them to miss deadlines. However, CPU
%performance is high. 
%\MODIFIEDISCA{On the other hand, when we use optimized values of {\it
%EmergentThreshold} (FRFCFS-DynOpt), the HWAs are prioritized enabling them to
%meet almost all their deadlines, but at the cost of CPU performance. Third,
%SQUASH achieves comparable performance to FRFCFS-Dyn-0.9 and 10.1\% better
%system performance than FRFCFS-DynOpt, while achieving a deadline-met ratio of
%100\%. 
%%This comes at the cost of 44.5 \% higher unfairness, compared to
%%FRFCFS-DynOpt, from prioritizing HWAs over memory-intensive applications. 
%We conclude that SQUASH achieves both high CPU performance and almost 100\% QoS for
%HWAs. In the next section, we present a breakdown of the benefits from the
%different components of SQUASH.}

\subsection{Performance Breakdown of SQUASH} \label{sec:intensity} 
\begin{figure}[h]
  \centering
    \includegraphics[scale=0.25, angle=270]{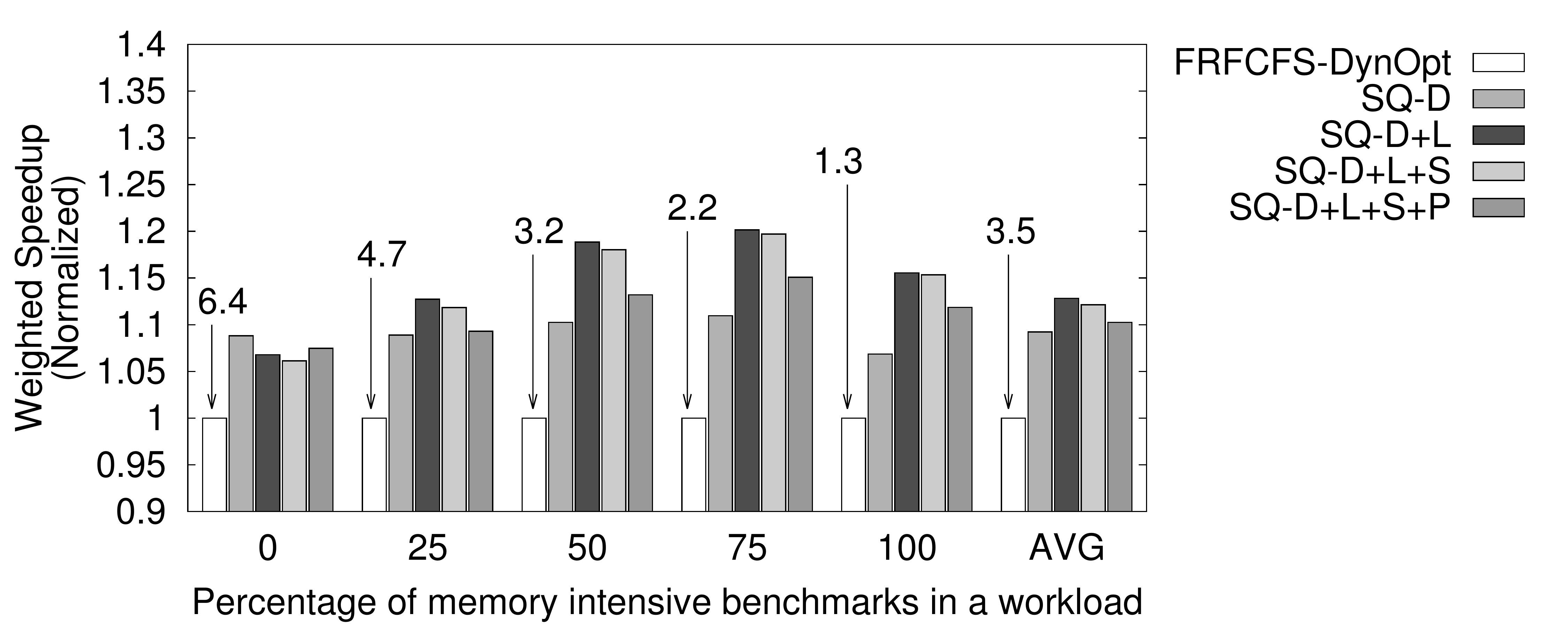}
%  \caption{System performance and fairness of SQUASH for different workload memory intensities}
%    \vspace{-2mm}
  \caption{SQUASH performance breakdown for different workload memory intensities}
  \label{plot:intensity_results}
\vspace{-3mm}
\end{figure}

\begin{figure}[h]
%  \vspace{-3mm}
  \centering
  \begin{minipage}{0.40\textwidth}
    \centering
    \includegraphics[scale=0.25, angle=270]{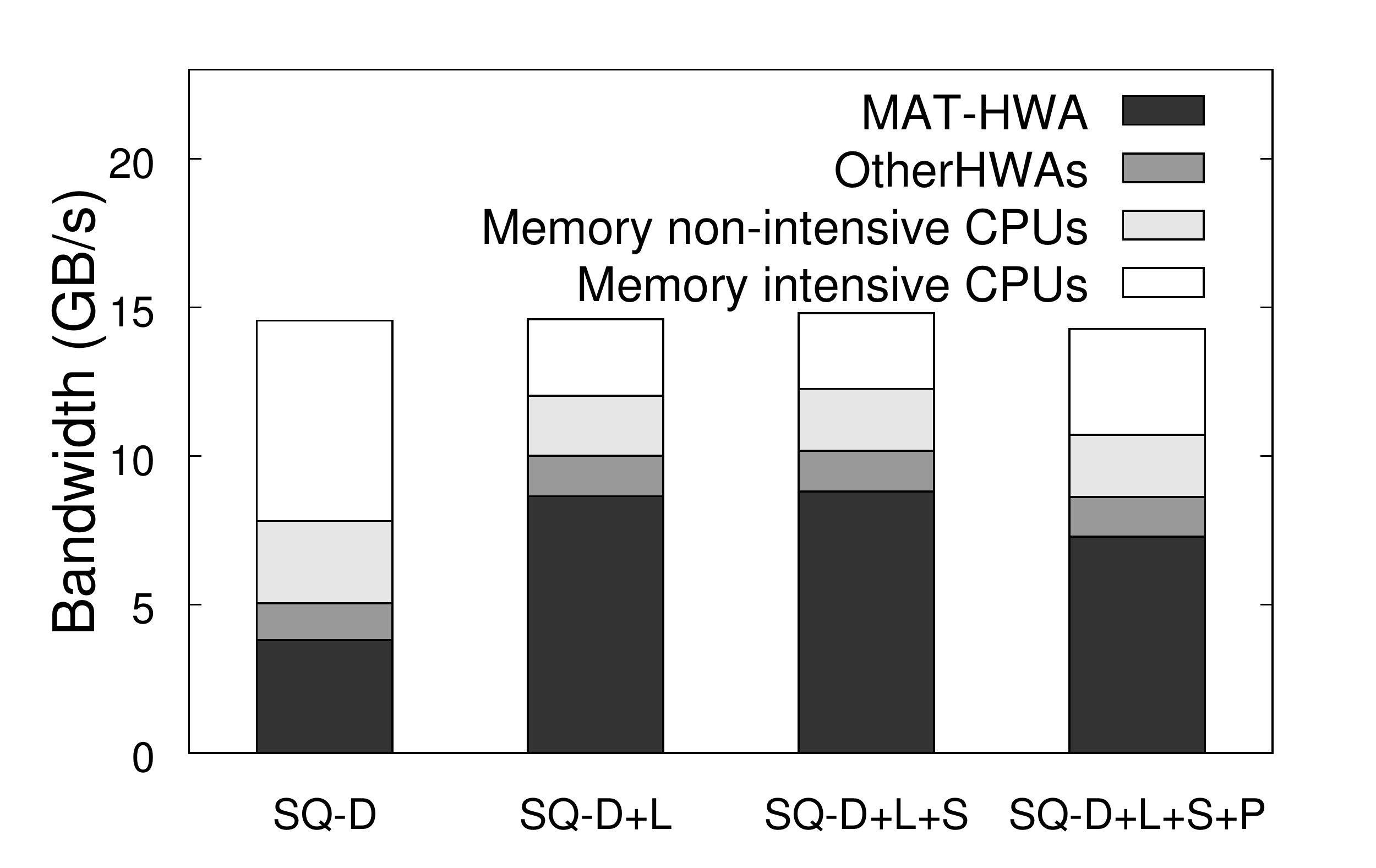}
  \end{minipage}
  \begin{minipage}{0.40\textwidth}
    \centering
    \includegraphics[scale=0.25, angle=270]{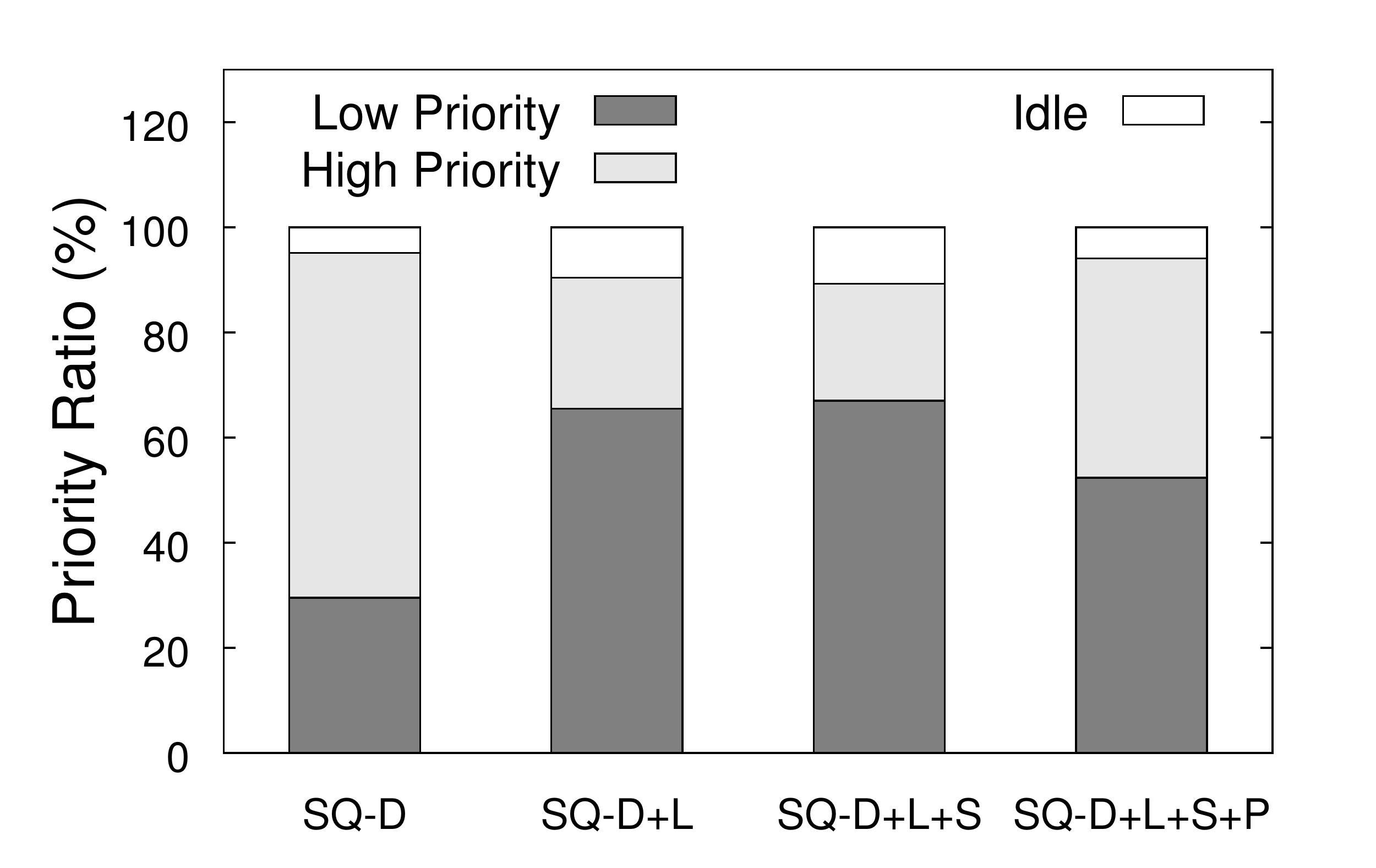}
  \end{minipage}
%  \vspace{-2mm}
  \caption{Bandwidth and priority ratio}
  \label{plot:bw_ratio_hwa}
  %\vspace{-4mm}
\end{figure}

In this section, we break down the performance benefits due to the different components
of SQUASH. Figure~\ref{plot:intensity_results} shows the system performance 
normalized to FRFCFS-DynOpt. The x-axis shows the memory intensities of the
workloads. The numbers above the bars of FRFCFS-DynOpt show the absolute values
for FRFCFS-DynOpt. 
We compare four different configurations of SQUASH over
FRFCFS-DynOpt: 1) SQ-D (distributed priority on top of TCM for CPU
applications), 2) SQ-D+L (SQ-D along with application-aware prioritization
between \ldphwas and memory-intensive CPU applications), 3) SQ-D+L+S (SQ-D+L
along with worst-case latency based prioritization for \sdphwas), and 4)
SQ-D+L+S+P (Complete SQUASH mechanism, SQ-D+L+S along with probabilistic prioritization between \ldphwas
and memory-intensive CPU applications). 
Table~\ref{tab:average_hwa_performance2} shows the deadline-met ratio for the
different mechanisms. Figure~\ref{plot:bw_ratio_hwa} shows the bandwidth
utilization of different categories of applications when the MAT-HWA has low
priority and the fraction of time the MAT-HWA is assigned different priorities.

% Original Longer Version
%In this section, we break down the performance benefits due to the different components
%of SQUASH. Figure~\ref{plot:intensity_results} shows the system performance and
%normalized to FRFCFS-DynOpt. The x-axis shows the memory intensities of the
%workloads. The numbers above the bars of FRFCFS-DynOpt show the absolute values
%for FRFCFS-DynOpt. We compare four different configurations of SQUASH over
%FRFCFS-DynOpt: 1) SQ-D (distributed priority on top of TCM for CPU
%applications), 2) SQ-D+L (SQ-D along with application-aware prioritization
%between \ldphwas and memory-intensive CPU applications), 3) SQ-D+L+S (SQ-D+L
%along with worst-case latency based prioritization for \sdphwas), and 4)
%SQ-D+L+S+P (SQ-D+L+S along with probabilistic prioritization between \ldphwas
%and memory-intensive CPU applications). The complete SQUASH mechanism that we
%show in Figure~\ref{plot:average_performance} is SQ-D+L+S+P.
%Table~\ref{tab:average_hwa_performance2} shows the deadline-met ratio for the
%different mechanisms. Figure~\ref{plot:bw_ratio_hwa} shows the bandwidth
%utilization of different categories of applications when the MAT-HWA has low
%priority and the fraction of time the MAT-HWA is assigned different priorities.

%\begin{figure}[!t]
%  \centering
%  \begin{minipage}{0.23\textwidth}
%    \centering
%    \includegraphics[scale=0.17, angle=270]{plots/intensity_results_ws}
%  \end{minipage}
%%  \begin{minipage}{0.32\textwidth}
%%    \centering
%%    \includegraphics[scale=0.22, angle=270]{plots/intensity_results_hs}
%%  \end{minipage}
%  \begin{minipage}{0.23\textwidth}
%    \centering
%    \includegraphics[scale=0.17, angle=270]{plots/intensity_results_ms}
%  \end{minipage}
%%  \caption{System performance and fairness of SQUASH for different workload memory intensities}
%  \caption{SQUASH performance breakdown for different workload memory intensities}
%  \label{plot:intensity_results}
%\end{figure}

\begin{table}[h]
\footnotesize
\centering
\setlength{\tabcolsep}{.60em}
    \begin{tabular}{lcccc}
      \toprule
%      \multirow{2}{*}{\textbf{\begin{minipage}{0.5in}Scheduling Algorithms\end{minipage}}} &
      \multirow{2}{*}{\textbf{Scheduling Algorithms}} &
      \multicolumn{4}{c}{\textbf{Deadline-Met Ratio (\%) / Frame Rate (fps)}} \\
       & HES & MAT & RSZ & DET \\
      \midrule

    FRFCFS-DynOpt & 100 / 30 & 99.997 / 29.72 & 100 / 30 & 99.99 / 25.5 \\ 
%    FRFCFS-Dyn0.3 & 99.996 / 27.84 & 97.13 / 26.26  & 100 / 30 & 100 / 30 \\
%    FRFCFS-Dist & 99.91 / 17.81 & 99.999 / 29.81 & 100 / 30 & 99.72 / 23.06 \\
    SQ-D & 99.999 / 29.875 &  100 / 30 & 100 / 30 & 99.88 / 21 \\
    SQ-D+L & 99.999 / 29.934 & 100 / 30 & 100 / 30 & 99.86 / 20.44 \\
    SQ-D+L+S & 100 / 30 & 100 / 30 & 100 / 30 & 100 / 30 \\
    SQ-D+L+S+P & 100 / 30 & 100 / 30 & 100 / 30 & 100 / 30 \\

      \bottomrule
    \end{tabular}
%\vspace{-1mm}
\caption{Deadline-met ratio and frame rate of HWAs for SQUASH components}
\label{tab:average_hwa_performance2}%
\vspace{-2mm}
\end{table}

%\vspace{-0.2in}

%\begin{figure}[ht!]
%\vspace{-2mm}

We draw four major observations. First, SQ-D improves performance over
FRFCFS-DynOpt by 9.2\%. However, this improvement comes at the cost of missed
deadlines for some HWAs (HES and DET), as shown in
Table~\ref{tab:average_hwa_performance2}. Second, introducing application-aware
prioritization between \ldphwas and memory-intensive CPU applications (SQ-D+L)
%enables better deadline-met ratios for \ldphwas such as MAT, while also
improves performance especially as the memory intensity increases (8.3\% maximum over SQ-D). 
This is because prioritizing HWAs over memory-intensive applications reduces 
the amount of time HWAs become urgent and interfere with memory-non-intensive CPU applications 
as shown in Figure~\ref{plot:bw_ratio_hwa}. However, the \sdphwas (HES and DET)
still miss some deadlines.

Third, SQ-D+L+S employs worst-case access latency based prioritization
for \sdphwas, enabling such HWAs to meet their deadlines, while still achieving
high performance. However, memory-intensive applications still experience high
slowdowns. Fourth, SQ-D+L+S+P tackles this problem by probabilistically
changing the prioritization order between memory-intensive CPU applications and
\ldphwas. This increases the bandwidth allocation of memory-intensive CPU
applications, as shown in Figure~\ref{plot:bw_ratio_hwa}. %Furthermore, it also
%reduces the amount of time the HWA is idle. 
The result is a 26\% average reduction in the maximum slowdown experienced by any application in a workload,
while degrading performance by only 1.7\% compared to SQ-D+L+S and achieving
100\% deadline-met ratio. We conclude that SQUASH is in achieving high CPU
performance, while also meeting the HWAs' deadlines.

\subsection{Impact of \emph{EmergentThreshold}}\label{sec:sweeping_th}

\begin{figure}[ht!]
  \centering
  \includegraphics[scale=0.25, angle=270]{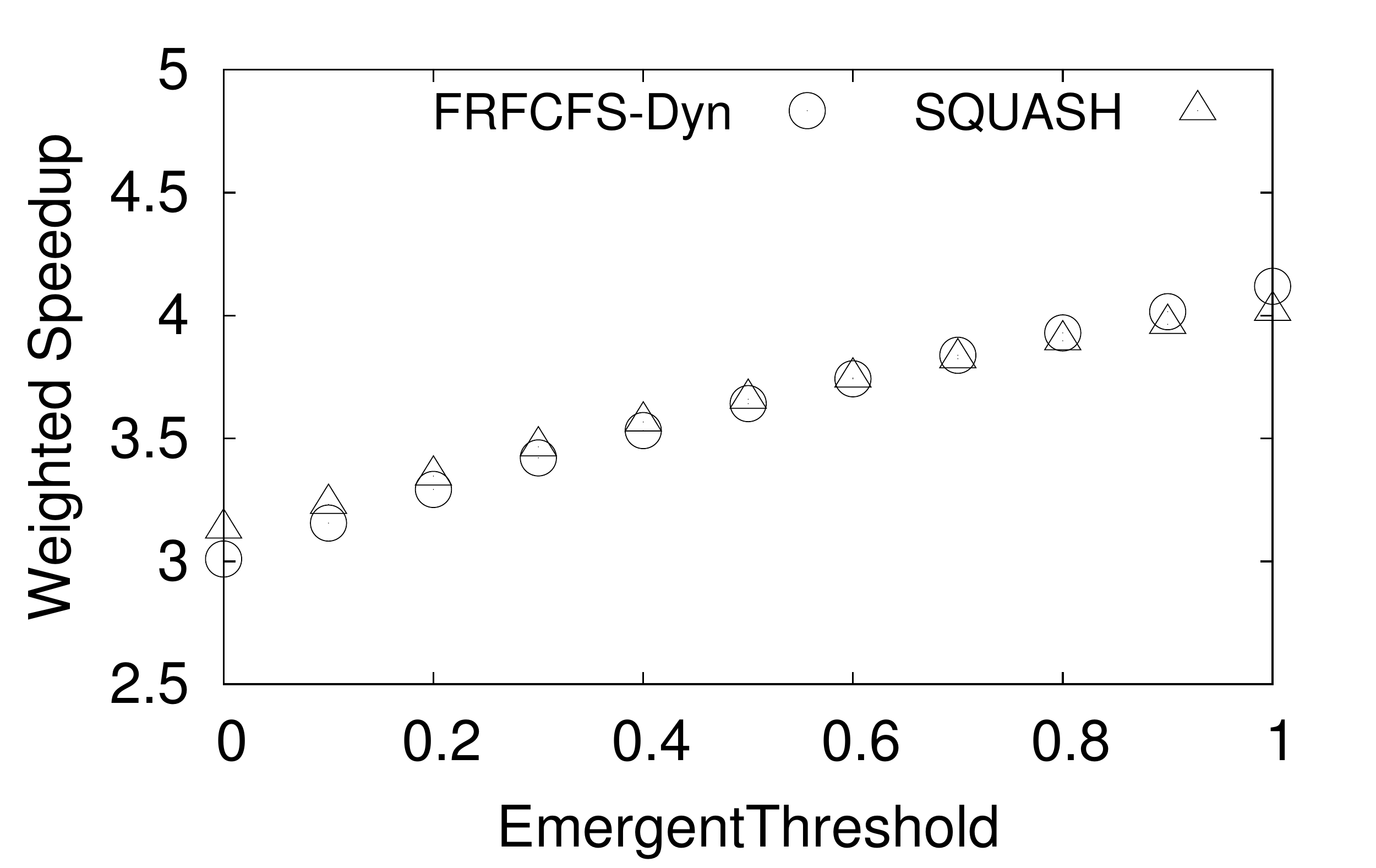}
%  \vspace{-2mm}
  \caption{Performance sensitivity to \emph{EmergentThreshold}}
  \label{plot:sweep_threshold}
\end{figure}

In this section, we study the impact of {\it EmergentThreshold} on
performance and deadline met ratio and the trade offs it enables.
Figure~\ref{plot:sweep_threshold} shows the average system
performance with FRFCFS-Dyn and SQUASH when sweeping {\it
EmergentThreshold} across all 80 workloads using both Config-A and
B. We employ the same {\it EmergentThreshold} value for all HWAs.
%\ifSQUEEZE
%Tables~\ref{tab:deadline-met-sweep} and
%\ref{tab:deadline-met-sweep-sq} show the deadline-met ratio of
%HWAs.
%\else
\TMP{Tables~\ref{tab:deadline-met-sweep} and
\ref{tab:deadline-met-sweep-sq} show the deadline-met ratio of
HWAs with FRFCFS-Dyn and SQUASH respectively. }
%\fi

%% Original Longer Version
%In this section, we study the impact of {\it EmergentThreshold} on
%performance and deadline met ratio and the trade offs it enables.
%Figure~\ref{plot:sweep_threshold} shows the average system
%performance with FRFCFS-Dyn and SQUASH when sweeping {\it
%EmergentThreshold} across all 80 workloads using both Config-A and
%B. For the purposes of this study, we employ the same one {\it
%EmergentThreshold} value for all HWAs.
%Tables~\ref{tab:deadline-met-sweep} and
%~\ref{tab:deadline-met-sweep-sq} show the deadline-met ratio of
%HWAs with FRFCFS-Dyn and SQUASH respectively. 

%\vspace{-5mm}
%\vspace{-9mm}
\vspace{-3mm}
\begin{table}[h!]
%\vspace{-5mm}
%\scriptsize
\footnotesize
\centering
\setlength{\tabcolsep}{.60em}
    \begin{tabular}{lcccccc}
      \toprule
      \multirow{3}{*}{\textbf{\begin{minipage}{0.5in}Emergent Threshold\end{minipage}}} &
      \multicolumn{6}{c}{\textbf{Deadline-Met Ratio (\%)}} \\
      & \multicolumn{2}{c}{\textbf{Config-A}} & \multicolumn{4}{c}{\textbf{Config-B}} \\
       & HES & MAT & HES & MAT & DET & RSZ \\
      \midrule
%      0&100&100&100&100&100&100 \\
      0-0.1&100&100&100&100&100&100 \\
      0.2&100&99.987&100&100&100&100 \\
      0.3&99.992&93.740&100&100&100&100 \\
      0.4&99.971&73.179&100&100&100&100 \\
      0.5&99.945&55.760&99.9996&99.751&99.997&100 \\
      0.6&99.905&44.691&99.989&94.697&99.960&100 \\
      0.7&99.875&38.097&99.957&86.366&99.733&100 \\
      0.8&99.831&34.098&99.906&74.690&99.004&99.886 \\
      0.9&99.487&31.385&99.319&60.641&97.149&97.977 \\
      1&96.653&27.320&95.798&33.449&88.425&55.773 \\
      \bottomrule
    \end{tabular}
\vspace{-2mm}
\caption{Deadline-met ratio of FRFCFS-Dyn}
\vspace{-2mm}
\label{tab:deadline-met-sweep}%
\end{table}

\begin{table}[h!]
%\vspace{-5mm}
\footnotesize
\centering
\setlength{\tabcolsep}{.60em}
    \begin{tabular}{lcccccc}
      \toprule
      \multirow{3}{*}{\textbf{\begin{minipage}{0.5in}Emergent Threshold\end{minipage}}} &
      \multicolumn{6}{c}{\textbf{Deadline-Met Ratio (\%)}} \\
      & \multicolumn{2}{c}{\textbf{Config-A}} & \multicolumn{4}{c}{\textbf{Config-B}} \\
       & HES & MAT & HES & MAT & DET & RSZ \\
      \midrule
      0-0.8&100&100&100&100&100&100 \\
      0.9&100&99.997&100&99.993&100&100 \\
      1.0&100&68.44&100&75.83&100&95.93 \\
      \bottomrule
    \end{tabular}
\vspace{-2mm}
\caption{Deadline-met ratio of SQUASH}
\label{tab:deadline-met-sweep-sq}%
\end{table}

We draw two major conclusions. First, there is a trade-off between system
performance and HWA deadline-met ratio, as the {\it EmergentThreshold} is
varied. As the {\it EmergentThreshold} increases, CPU performance improves
at the cost of increase in deadline-met ratio.
Second, for a given value of {\it EmergentThreshold}, SQUASH achieves
significantly higher deadline-met ratio than FRFCFS-Dyn, while achieving similar
CPU performance, because of distributed priority and application-aware scheduling.
Specifically, SQUASH meets all deadlines with an {\it EmergentThreshold} of 0.8,
for Config-A, whereas FRFCFS-Dyn needs an {\it EmergentThreshold} of 0.1 to meet
all deadlines. Furthermore, SQUASH-0.8 achieves 23.5\% higher performance
than FRFCFS-Dyn-0.1. Based on these observations, we conclude that SQUASH is
effective in achieving both high CPU performance and QoS for HWAs.

\subsection{Impact of \emph{ClusterFactor}}\label{sec:sweeping_th}

\begin{figure}[ht!]
  \centering
  \includegraphics[scale=0.25, angle=270]{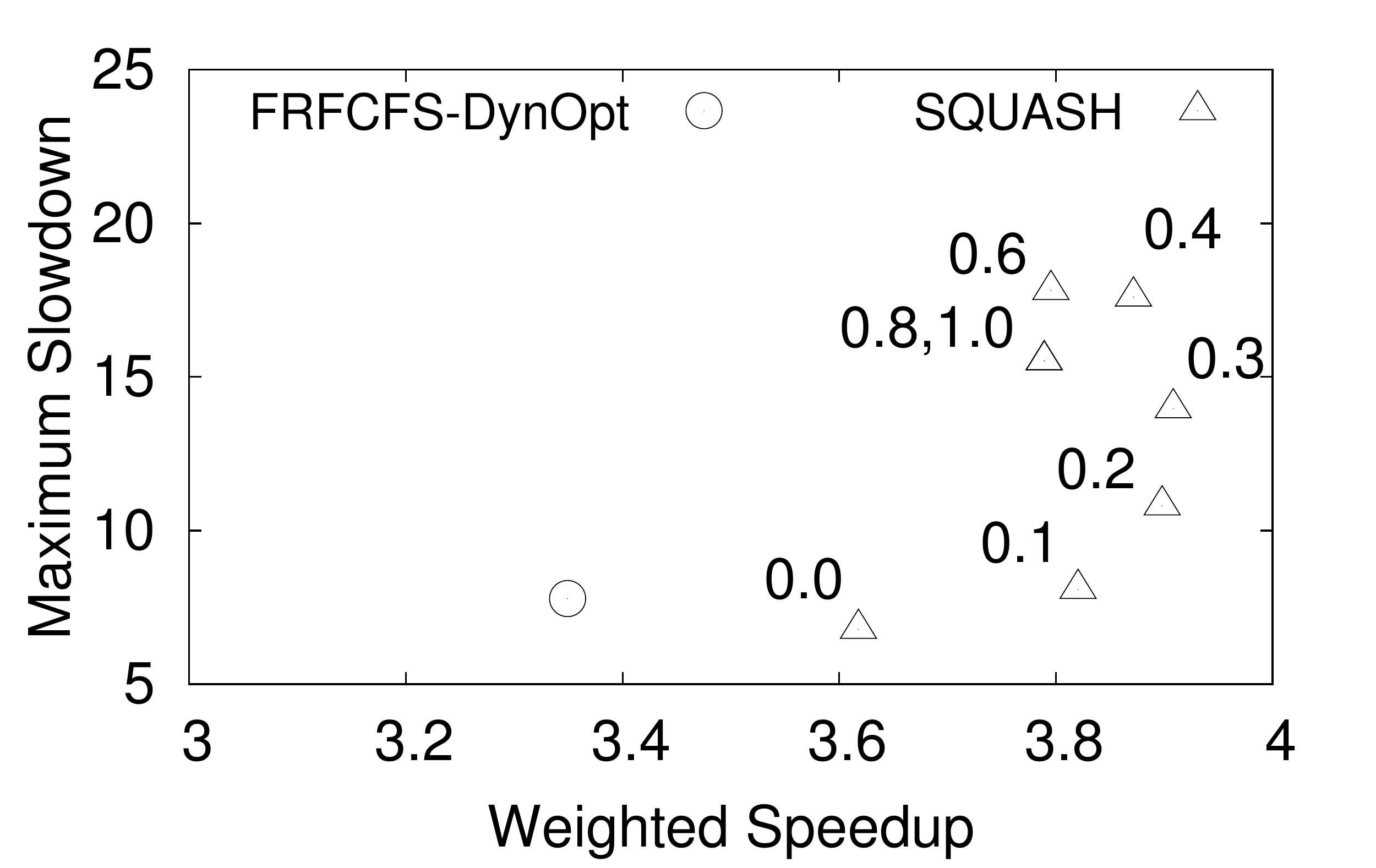}
%  \vspace{-2mm}
  \caption{Performance sensitivity to \emph{ClusterFactor}}
  \label{plot:sweep_tcmthreshold}
\end{figure}

We study the impact of the \emph{ClusterFactor} used to determine what fraction of
total memory bandwidth is allocated to memory-non-intensive CPU applications.
Figure~\ref{plot:sweep_tcmthreshold} shows average CPU performance and
fairness with FRFCFS-DynOpt and SQUASH across 80 workloads using Config-A.
For SQUASH, we sweep the \emph{ClusterFactor} from 0 to 1.0. All HWAs' deadlines are
met for all values of the \emph{ClusterFactor} for SQUASH.

%\vspace{-5mm}
%\vspace{-5mm}

We draw three major conclusions. First, there is a trade-off between performance
and fairness, as the \emph{ClusterFactor} is varied. As
the \emph{ClusterFactor} increases, CPU performance improves, but fairness degrades.
This is because more CPU applications are classified and prioritized
as memory-non-intensive at the cost of degrading the
performance of some memory-intensive applications.
Second, \emph{ClusterFactor} is an effective knob for trading off CPU
performance and fairness. For example, in our main evaluations, we optimize for
performance and pick a \emph{ClusterFactor} of 0.2. Instead, if we want to optimize
for fairness, we could pick a \emph{ClusterFactor} of 0.1, which still improves
performance by 14\%, compared to FRFCFS-DynOpt, while degrading fairness by only
3.8\% (for Config-A). Third, regardless of the \emph{ClusterFactor}, SQUASH is able to
meet all HWAs' deadlines (not shown), since it monitors and assigns enough priority
to HWAs based on their progress.

\subsection{Effect of HWAs' Memory Intensity}

\ifSQUEEZE
We study the impact of HWAs' memory intensity 
on a system with 2 MAT-HWAs and 2 HES-HWAs. 
\else
We study the impact of HWAs' memory intensity on performance and
deadline-met-ratio. We perform these evaluations on a system with 4 HWAs: 2
MAT-HWAs and 2 HES-HWAs. 
\fi
We vary the memory intensity of the HWAs by varying
their parameters in Table~\ref{tab:hwa_env}. 
%(We do not show these plots due to space constraints). 
As the HWAs' memory intensity increases, CPU performance improvement with SQUASH
increases (30.6\% maximum) while meeting almost all deadlines (99.99\%), when
using an \emph{EmergentThreshold} of 0.8. This is because as the memory intensity
of HWAs increases, they cause more interference to CPU applications. SQUASH is
effective in mitigating this interference.

\subsection{Evaluation on systems with GPUs}
\begin{figure}[ht!]
%  \vspace{-8mm}
  \centering
%  \flushleft
  \begin{minipage}{0.40\textwidth}
%    \flushleft
    \centering
    \includegraphics[scale=0.25, angle=270]{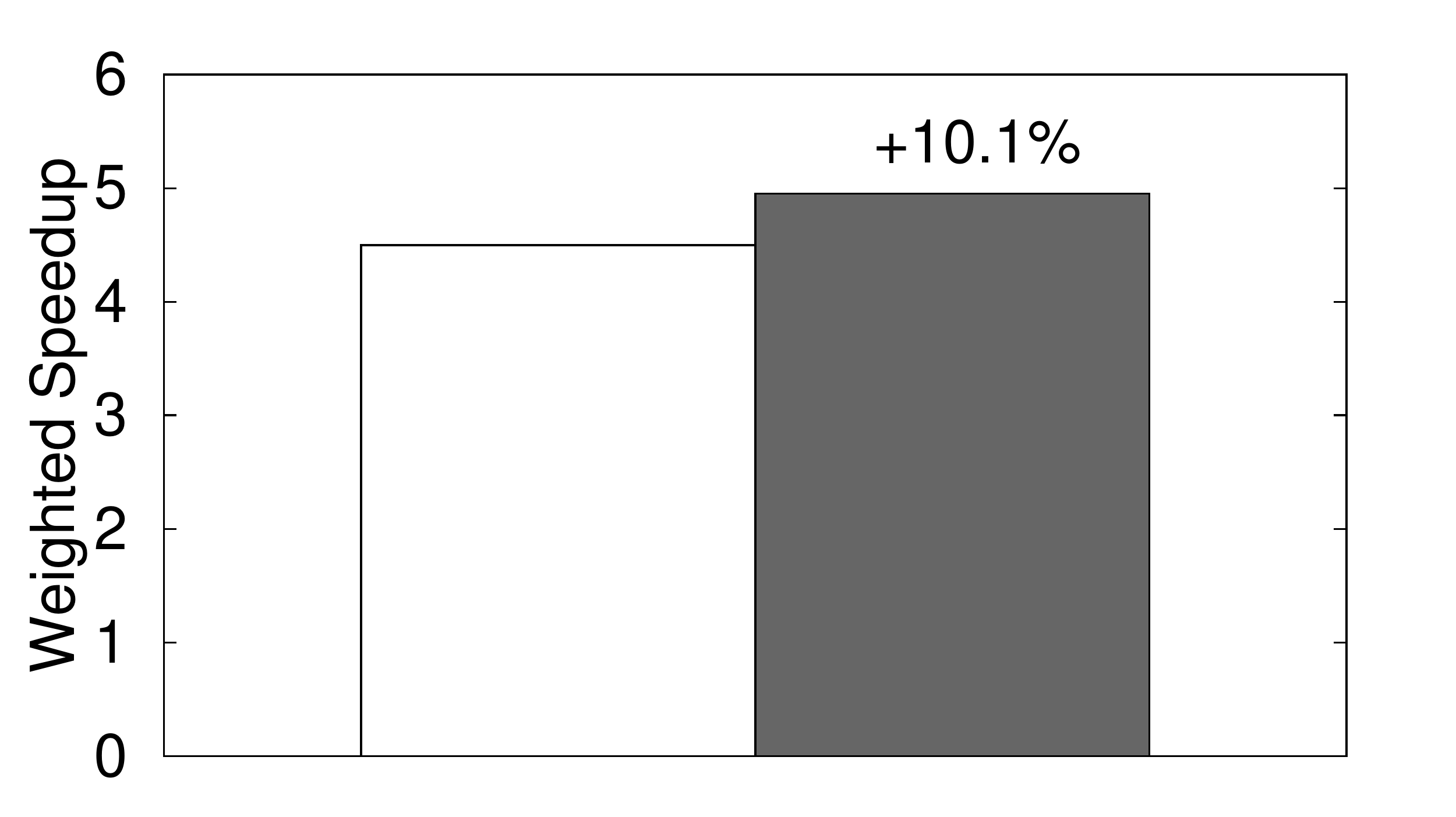}
  \end{minipage}
  \begin{minipage}{0.40\textwidth}
    \centering
%    \flushleft
    \includegraphics[scale=0.25, angle=270]{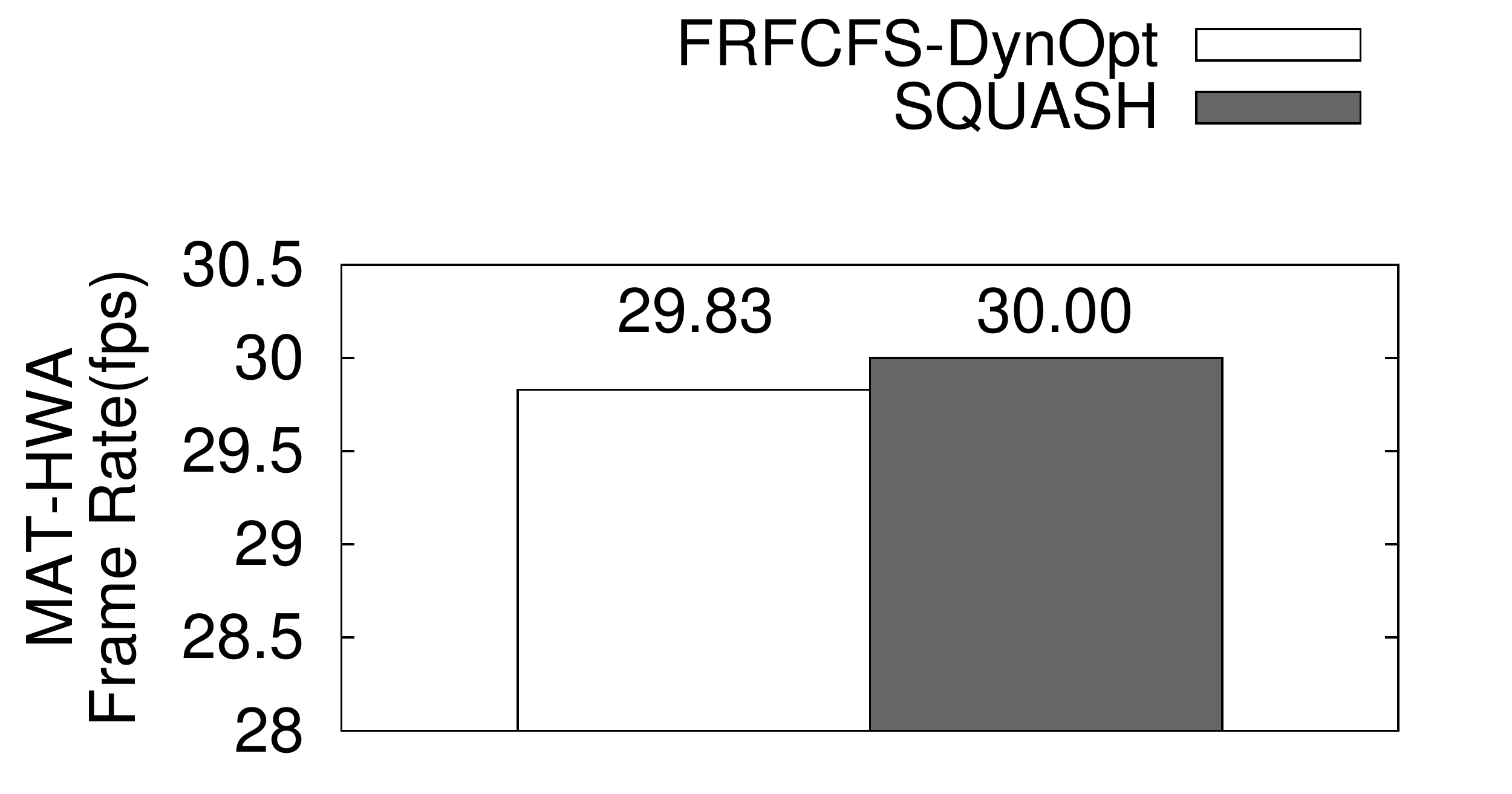}
  \end{minipage}
%  \vspace{-3mm}
  \caption{System performance and frame rate}
  \label{plot:gpu_average_performance}
\end{figure}

Figure~\ref{plot:gpu_average_performance} shows the average CPU performance
and frame rate of the MAT-HWA across 30 workloads on a CPU-GPU-HWA
system. The other HWAs and the GPU meet all deadlines with all schedulers.
%Table~\ref{tab:gpu_hwa_performance} shows the deadline-met ratio and frame rate
%of GPU and FixedHWAs. We do not show HWAs that achieve 100\% deadline-met with
%all schedulers.
For FRFCFS-Dyn, we use an {\it EmergentThreshold} of 0.9 for the GPU and the
threshold values shown in Table~\ref{tab:emergent_threshold} for the other HWAs.
For SQUASH, we use an {\it EmergentThreshold} of 0.9 for both the GPU and other HWAs.

% Original Longer Version
%For FRFCFS-Dyn, we use an {\it EmergentThreshold} of 0.9 for the GPU and the
%same threshold values as the CPU-HWA system for the other HWAs(shown in
%Table~\ref{tab:emergent_threshold}). For SQUASH, we use an {\it
%EmergentThreshold} of 0.9 for both the GPU and other HWAs.

%\vspace{-5mm}
%\begin{figure}[ht!]
%  \centering
%  \includegraphics[scale=0.17, angle=270]{plots/main_gpuhwa_unnorm_ws}
%  \vspace{-2mm}
%  \caption{System performance}
%  \label{plot:gpu_average_performance}
%\end{figure}
%\vspace{-8mm}

%\vspace{-1mm}

%\input{sections/deadline_results_gpu}
%\vspace{-5mm}

SQUASH achieves 10.1\% higher CPU performance than FRFCFS-Dyn, while also
achieving a higher frame rate for the MAT-HWA than FRFCFS-Dyn. SQUASH's
distributed priority and application-aware scheduling schemes enable higher
system performance, while ensuring QoS for the HWAs \emph{and} the GPU.  We also
evaluate a CPU-GPU system. SQUASH improves CPU performance by 2\% over
FRFCFS-Dyn, while meeting all deadlines, where FRFCFS-Dyn misses a deadline. We
conclude that SQUASH is effective in achieving high system performance and QoS
in systems with GPUs as well.

\subsection{Sensitivity to System Parameters}
\subsubsection{Number of Channels.} 
\begin{figure}[ht!]
  \centering
  \begin{minipage}{0.40\textwidth}
    \centering
    \includegraphics[scale=0.25, angle=270]{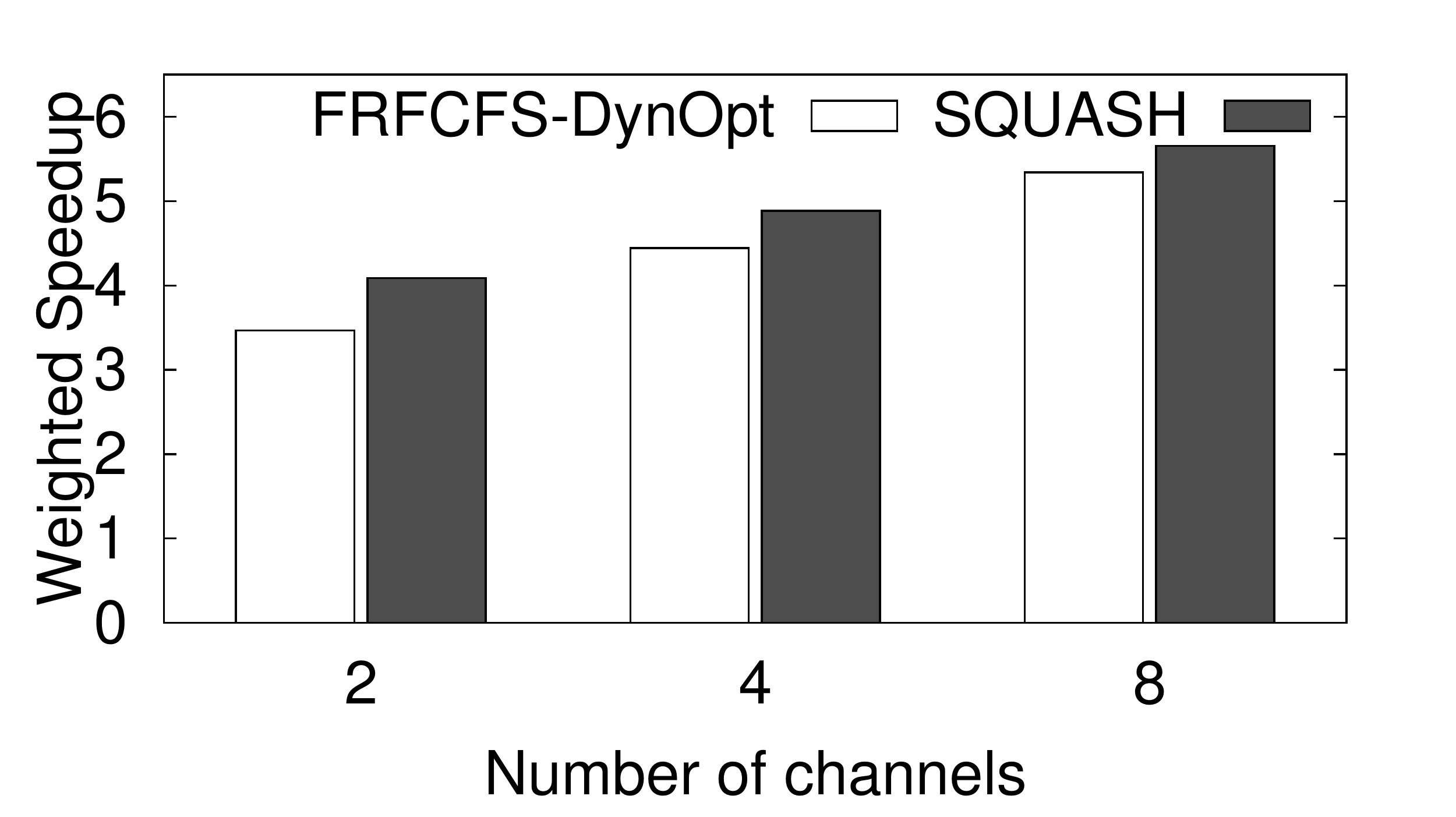}
  \end{minipage}
%%  \begin{minipage}{0.24\textwidth}
%%    \centering
%%    \includegraphics[scale=0.16, angle=270]{plots/ch_num_hs}
%%  \end{minipage}
  \begin{minipage}{0.40\textwidth}
    \centering
    \includegraphics[scale=0.25, angle=270]{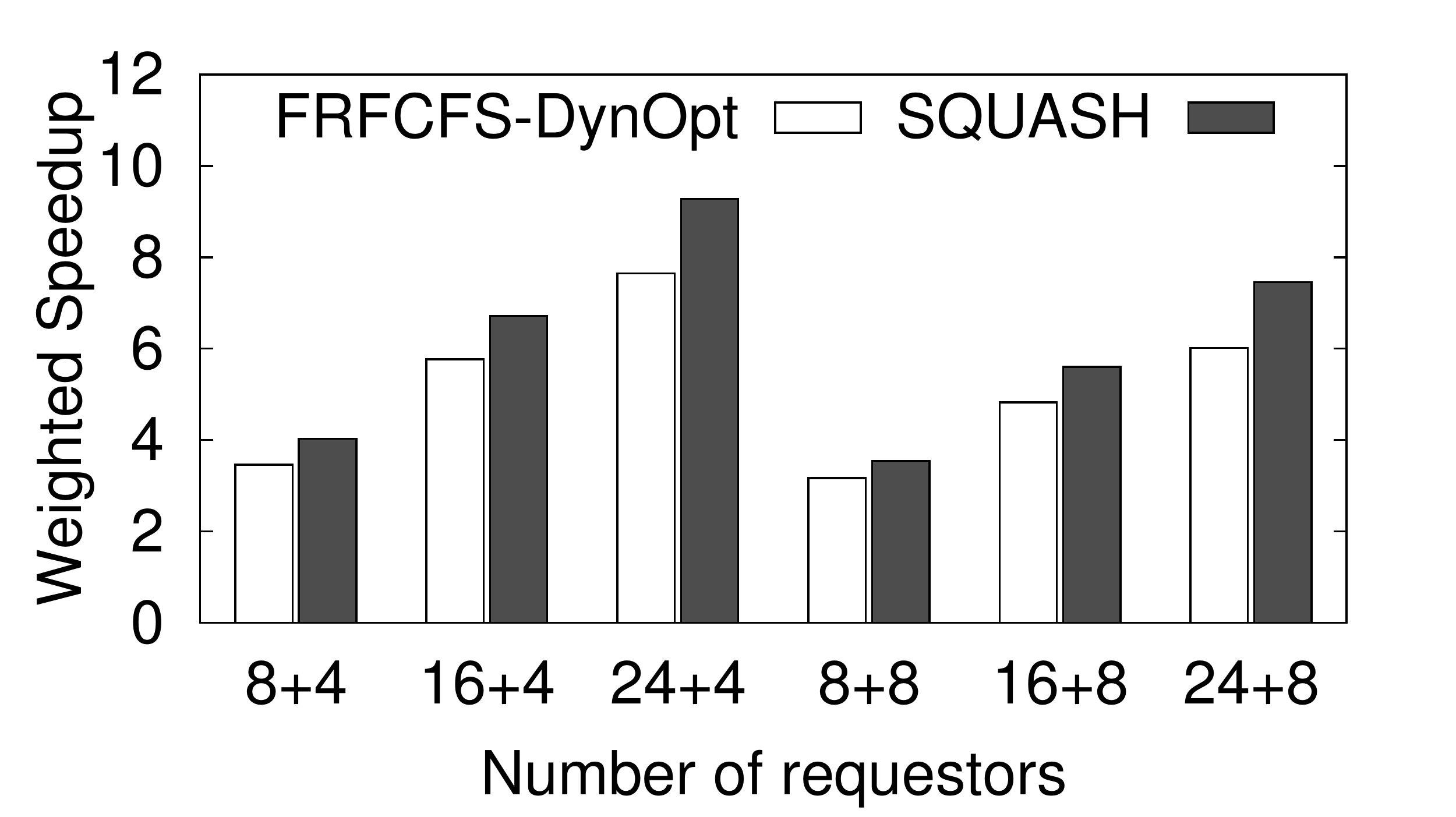}
  \end{minipage}
%  \vspace{-2mm}
  \caption{Performance sensitivity to system parameters}
  \label{plot:ch_num_results}
\end{figure}

\ifSQUEEZE
Figure~\ref{plot:ch_num_results} (left) shows the system performance
with different number of channels across 25 workloads
(executing 90M cycles) using HWA Config-A (all other parameters are the
same as baseline). 
All HWAs meet all deadlines with all schedulers as there is ample bandwidth. The key
conclusion is that as the number of channels decreases, memory contention increases, 
resulting in increased benefits from SQUASH.
\else
The left of figure~\ref{plot:ch_num_results} shows the system performance of
FRFCFS-DynOpt and SQUASH with different number of channels across 25 workloads
(executing 90M cycles) using HWA Config-A (all other parameters are the
same as baseline). 
All HWAs meet all deadlines with all schedulers. The key
conclusion from this analysis is that as the number of
channels decreases, memory contention increases, resulting in increased benefits
from SQUASH.
\fi
%Figures~\ref{plot:ch_num_results} and
%\ref{plot:channel_num_ratio} show the system performance, fairness, and
%deadline-met ratio of FRFCFS-Dyn0.3, \noprob, and SQUASH with different number of
%channels across 25 workloads using HWA Config-A, when all other parameters are
%the same as baseline. 

%Two observations are in order. First, as the number of
%channels decreases, memory contention increases, resulting in increased benefits
%from \noprob and SQUASH. Second, SQUASH has both higher system performance and
%fairness compared to \noprob as the number of channels increases.
%This is because memory-intensive applications probabilistically receive higher
%bandwidth than LDP-HWAs without severely degrading LDP-HWAs' progress.

%This is
%because non-urgent HWAs receive enough memory bandwidth to make sufficient
%progress to meet their deadlines, even when they are prioritized below
%memory-intensive applications. SQUASH enables this better by virtue of
%probabilistically switching priorities between \ldphwas and
%memory-intensive CPU applications.

%\vspace{-5mm}
%\begin{figure}[ht!]
% \centering
% \includegraphics[scale=0.16, angle=270]{plots/ch_num_ws}
%  \caption{Performance sensitivity to channel count}
%  \label{plot:ch_num_results}
%\end{figure}

%\vspace{-5mm}
%\vspace{-4mm}

%\vspace{-5mm}
%\begin{figure}[ht!]
%  \flushleft
%  \begin{minipage}{0.23\textwidth}
%    \flushleft
%    \includegraphics[scale=0.16, angle=270]{plots/ch_num_ws}
%  \end{minipage}
%%  \begin{minipage}{0.24\textwidth}
%%    \centering
%%    \includegraphics[scale=0.16, angle=270]{plots/ch_num_hs}
%%  \end{minipage}
%  \begin{minipage}{0.23\textwidth}
%    \flushleft
%    \includegraphics[scale=0.16, angle=270]{plots/ch_num_ms}
%  \end{minipage}
%  \vspace{-1mm}
%  \caption{Performance sensitivity to channel count}
%  \label{plot:ch_num_results}
%\end{figure}

%\vspace{-9mm}
%\begin{figure}[ht!]
%  \flushleft
%  \begin{minipage}{0.23\textwidth}
%    \flushleft
%    \includegraphics[scale=0.16, angle=270]{plots/ch_num_methes}
%  \end{minipage}
%  \begin{minipage}{0.23\textwidth}
%    \flushleft
%    \includegraphics[scale=0.16, angle=270]{plots/ch_num_metmat}
%  \end{minipage}
%  \vspace{-1mm}
%  \caption{Deadline-met ratio sensitivity to channel count}
%  \label{plot:channel_num_ratio}
%\end{figure}
%\vspace{-4mm}

\subsubsection{Number of Cores.}

\begin{figure}[ht!]
  \centering
  \begin{minipage}{0.40\textwidth}
    \centering
    \includegraphics[scale=0.25, angle=270]{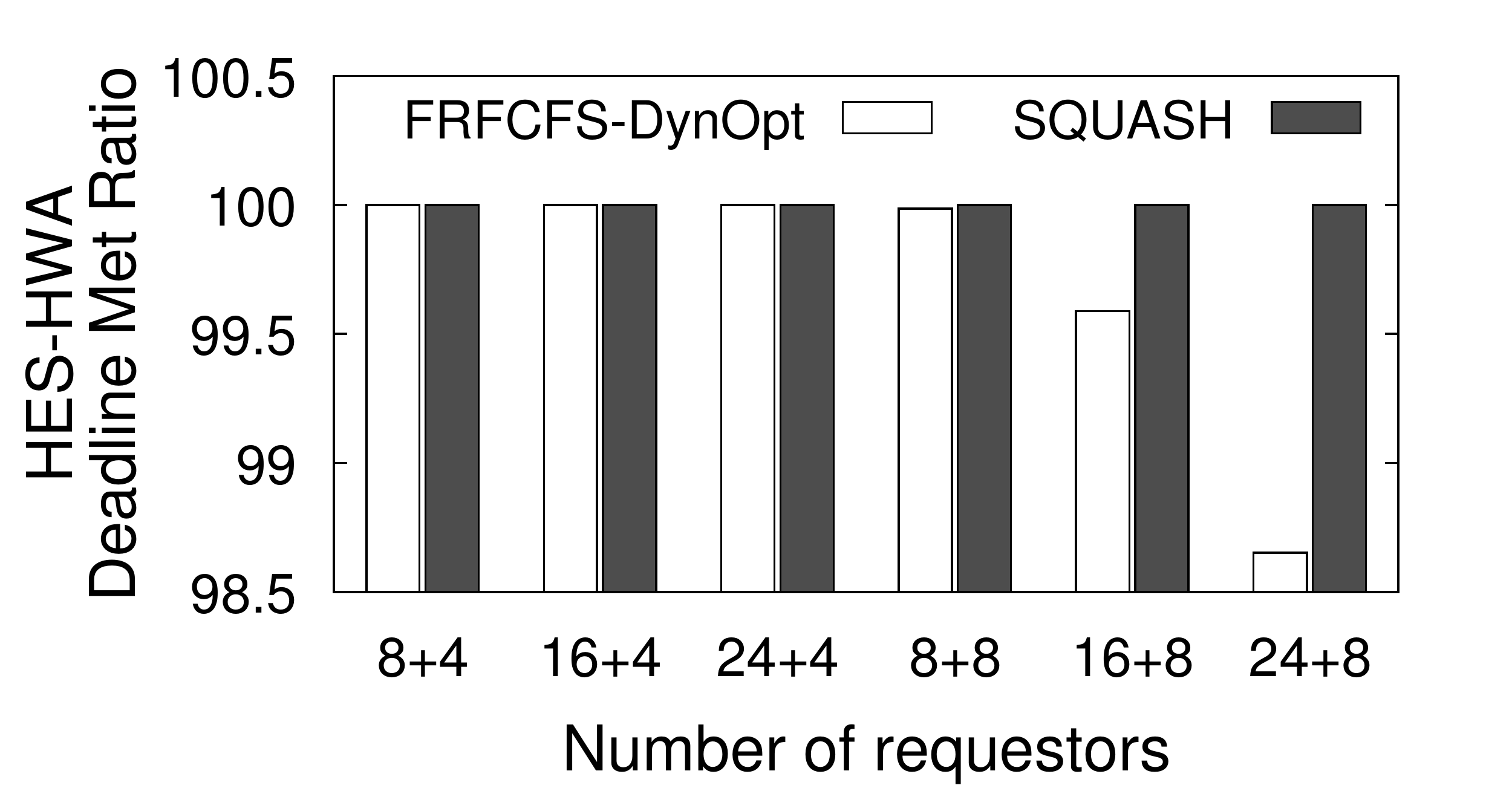}
  \end{minipage}
  \begin{minipage}{0.40\textwidth}
    \centering
    \includegraphics[scale=0.25, angle=270]{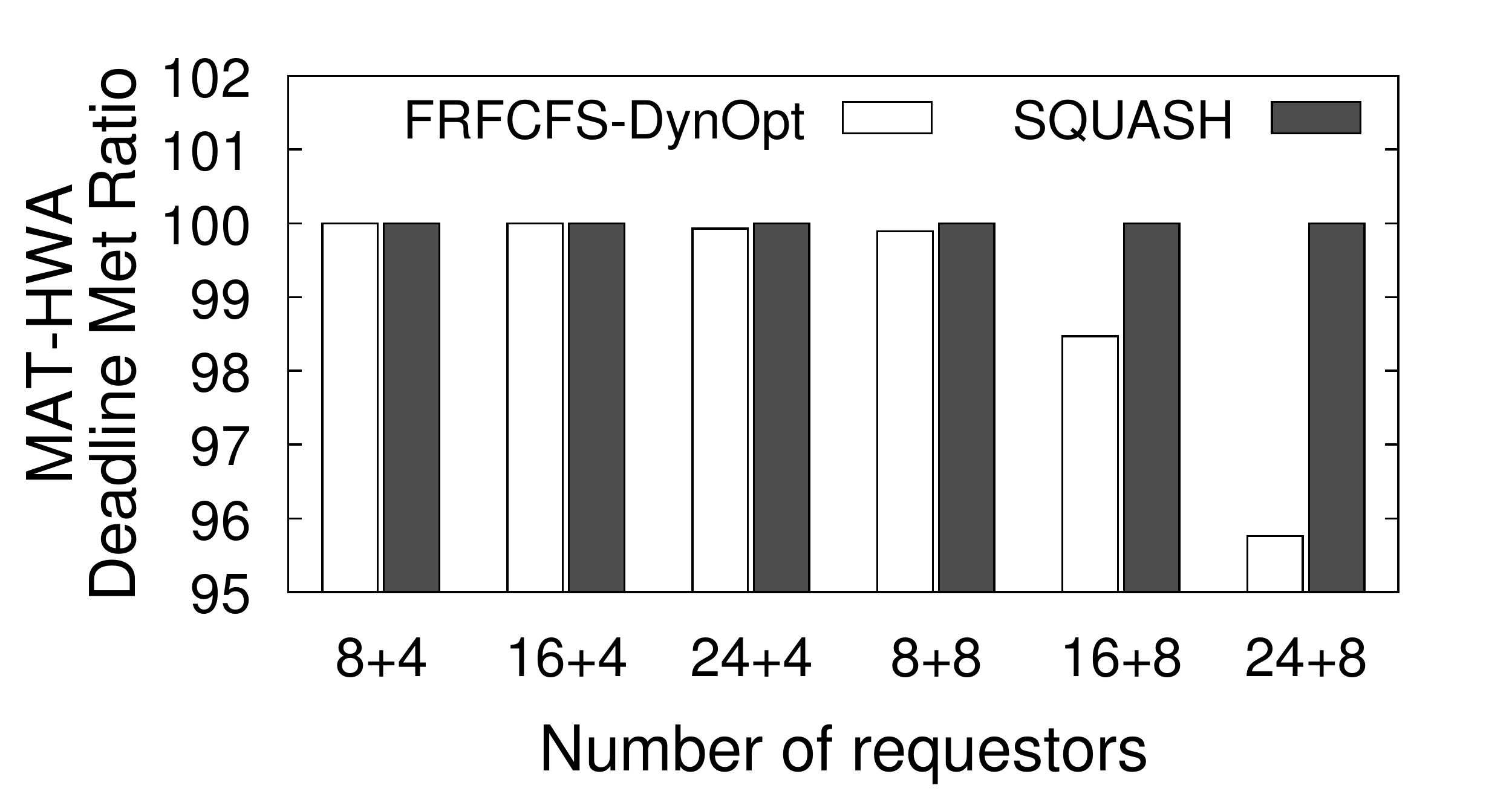}
  \end{minipage}
%  \vspace{-2mm}
  \caption{Deadline-met ratio sensitivity to core count}
  \label{plot:core_num_ratio}
\end{figure}

Figures~\ref{plot:ch_num_results} (right) and \ref{plot:core_num_ratio} show the same
performance metrics for the same schedulers as the previous section when using
different number of CPU cores (from 8 to 24) and HWAs\footnote{The 4-HWA
configuration is the same as Config-A. The 8-HWA configuration consists of
IMG-HWA x2, MAT-HWA(10) x1, MAT-HWA(20) x1, HES-HWA(32) x1, HES-HWA(128) x1,
RSZ-HWA x1, and DET-HWA x1. } (4 or 8). We draw three conclusions. First, SQUASH
\emph{always} improves CPU performance over FRFCFS-DynOpt. Second, as the number
of requestors increases, there is more contention between HWAs and CPU
applications, providing more opportunity for SQUASH, which achieves greater
performance improvement (24.0\% maximum).
%However, unfairness increases as the number of requestors increases. This is because non-urgent
%HWAs consume more memory bandwidth than memory-intensive applications. Second,
%SQUASH mitigates unfairness of \noprob by probabilistically switching
%priorities between HWAs and memory-intensive applications. 
Finally, SQUASH meets all deadlines for all HWAs.

%\vspace{-4mm}
%\vspace{-3mm}

\subsubsection{\emph{Scheduling Unit} and \emph{Switching Unit}.}
\ifSQUEEZE
We sweep the \textit{SchedulingUnit} (Section~\ref{sec:longdeadline}) from 1000 to 5000 cycles 
and \textit{SwitchingUnit} (Section~\ref{sec:prob-switching}) from 500 to 2000 cycles
(\textit{SwitchingUnit} $<$ \textit{SchedulingUnit}). 
We observe two trends.
First, as the \textit{SchedulingUnit} increases, system performance decreases
because once a HWA is classified as urgent, it interferes with CPU cores for
a longer time.
Second, a smaller \textit{SwitchingUnit} provides better fairness, since fine-grained 
switching of the probability $Pb$
enables memory-intensive applications to have higher priority for longer time
periods. Based on these observations, we empirically pick a \textit{SchedulingUnit} of 1000
cycles and \textit{SwitchingUnit} of 500 cycles.
\else
We sweep the \textit{SchedulingUnit}(Section~\ref{sec:longdeadline}) and
\textit{SwitchingUnit}(Section~\ref{sec:prob-switching}) across a wide range of
values for SQUASH. We evaluate values from 1000 to 5000 cycles for
\textit{SchedulingUnit} and 500 to 2000 cycles for \textit{SwitchingUnit}
(\textit{SwitchingUnit} $<$ \textit{SchedulingUnit}). 
We observe two trends.
First, as the \textit{SchedulingUnit} increases, system performance decreases
because once a HWA is classified as urgent, it interferes with CPU cores for
longer time.
%This is because, once a HWA is classified as urgent, it interferes with CPU
%cores for longer times with a longer \textit{SchedulingUnit}.
Second, a smaller \textit{SwitchingUnit} provides better fairness for a given
\textit{SchedulingUnit}, since fine-grained switching of the probability $Pb$
enables memory-intensive applications to have higher priority for longer time
periods. Based on these observations, we empirically pick a \textit{SchedulingUnit} of 1000
cycles and \textit{SwitchingUnit} of 500 cycles.
\fi

%\begin{figure}[ht!]
%  \flushleft
%  \begin{minipage}{0.23\textwidth}
%%    \centering
%    \flushleft
%    \includegraphics[scale=0.17, angle=270]{plots/scheduling_unit_performance}
%  \end{minipage}
%%  \begin{minipage}{0.24\textwidth}
%%    \centering
%%    \includegraphics[scale=0.16, angle=270]{plots/scheduling_unit_hs}
%%  \end{minipage}
%%  \begin{minipage}{0.24\textwidth}
%%    \centering
%%    \includegraphics[scale=0.16, angle=270]{plots/scheduling_unit_ms}
%%  \end{minipage}
%  \begin{minipage}{0.14\textwidth}
%    \flushleft
%    \includegraphics[scale=0.15, angle=270]{plots/scheduling_unit_avg_met}
%  \end{minipage}
%  \caption{Sensitivity to scheduling unit and switching unit}
%  \label{plot:schedulingunit_results}
%\end{figure}

\section{Related Work}

\noindent\textbf{Memory scheduling.} We have already compared
SQUASH both qualitatively and quantitatively to the state-of-the-art QoS-aware
memory scheduler for CPU-GPU systems proposed by Jeong et
al.~\cite{schedulingCPUGPU}. When this scheduler is adapted to the CPU-HWA context,
SQUASH outperforms it in terms of both system performance and deadline-met ratio.

Ausavarungnirun et al.~\cite{sms} propose Staged Memory Scheduling (SMS) to
improve system performance and fairness in a heterogeneous system
CPU-GPU system. Unlike SQUASH, SMS does not explicitly attempt to provide QoS to the GPU and aims
to only optimize overall performance and fairness.

Most other previously proposed memory
schedulers (e.g., ~\cite{frfcfs,stfm,parbs,fqm,atlas,tcm,mise,Hur_micro2004} have been designed with
the goal of improving system performance and fairness in CPU-only multicore
systems. These works do not consider the memory access characteristics and needs
of other requestors such as HWAs. In contrast, SQUASH is specifically designed to provide high system
performance and QoS in heterogeneous systems with CPU cores and HWAs.

Lee et al.~\cite{Qware} propose a quality-aware memory controller that aims to
satisfy latency and bandwidth requirements of different requestors, in a
best-effort manner. Latency-sensitive requestors are always given higher
priority over bandwidth-sensitive requestors. Hence, it might not be possible to
meet potential deadline requirements of bandwidth-sensitive requestors with
such a mechanism.

%Lee et al.\cite{Qware} propose a quality-aware memory controller that categorizes
%requestors into latency-sensitive, bandwidth-sensitive and don't-care channels.
%The latency-sensitive channel is
%Because latency-sensitive channels is always prior to bandwidth-sensitive channels,
%system designers have to consider bandwidth allocation carefully to meet the deadlines
%of bandwidth-sensitive HWAs.

Other previous works~\cite{predator,PRET,analyzable,timepred,beyond,Hyouseung_rtas2014} have
proposed to build memory controllers that provide support to guarantee
real-time access latency constraints for each master. The PRET DRAM
Controller~\cite{PRET} partitions DRAM into multiple resources that are
accessed in a periodic pipeline fashion. Wu et al.~\cite{timepred} propose to
strictly prioritize real-time threads over non real-time threads. Macian et
al.~\cite{beyond} bound the maximum access latency by scheduling in a
round-robin manner. Other works~\cite{predator,analyzable} group a series of accesses
to all banks and schedule at the group unit. All these works aim to bound
the worst case latency by scheduling requests in a fixed predictable order. As a result,
they waste significant amount of memory bandwidth and do not achieve high
system performance.

\noindent\textbf{Source throttling.} Memguard~\cite{memguard} guarantees worst
case bandwidth to each core by regulating the number of injected requests from
each core. Ebrahimi et al.~\cite{fst} propose to limit the number of memory requests of
requestors to improve fairness in CPU-only systems.
%The remaining bandwidth is used in a best-effort manner.
Other previous works~\cite{armwhite,schedulingCPUGPU} propose to throttle the
number of outstanding GPU requests for CPU-GPU systems, in order to mitigate
interference to CPU applications. These schemes are complementary to the memory
scheduling approach taken by SQUASH and can be employed in conjunction with
SQUASH to achieve better interference mitigation.

\noindent\textbf{Memory channel/bank partitioning.} Previous
works~\cite{mcp,bank-part,pact-bank-part} propose to mitigate interference by
mapping data of applications that significantly interfere with each other to
different channels/banks. Our memory scheduling approach can be combined with a
channel/bank partitioning approach to achieve higher system performance and QoS
for HWAs.

\section{Conclusion} We introduce a simple QoS-aware high-performance memory
scheduler for heterogeneous systems with hardware accelerators, SQUASH, with the
goal of enabling hardware accelerators (HWAs) to meet their deadlines while
providing high CPU performance. Our experimental evaluations across a wide variety of
workloads and systems show that SQUASH meets HWAs' deadlines and improves their
frame rates while also greatly improving CPU performance, compared to the
state-of-the-art techniques. We conclude that SQUASH can be an efficient and effective
memory scheduling substrate for current and future heterogeneous SoCs, which
will require increasingly more predictable and at the same time high performance
memory systems.

\newpage
\bibliographystyle{ieee}
{\small
%\bibliography{paper}
\bibliography{references}

\begin{thebibliography}{10}\itemsep=-1pt

\bibitem{nas}
{\em {NAS Parallel Benchmark Suite}}.
\newblock {http://www.nas.nasa.gov/publications/npb.html}.

\bibitem{spec2006}
{\em {SPEC CPU2006}}.
\newblock {http://www.spec.org/spec2006}.

\bibitem{tpc}
{\em {Transaction Processing Performance Council}}.
\newblock \url{http://www.tpc.org/}.

\bibitem{HWA_face}
L.~Acasandrei and A.~Barriga.
\newblock {AMBA bus hardware accelerator IP for Viola-Jones face detection}.
\newblock {\em {Computers Digital Techniques, IET}}, 7(5), September 2013.

\bibitem{amd-radeon}
{Advanced Micro Devices}.
\newblock {AMD Radeon HD 5870 Graphics}.

\bibitem{predator}
B.~Akesson, K.~Goossens, and M.~Ringhofer.
\newblock {Predator: A Predictable SDRAM Memory Controller}.
\newblock In {\em CODES+ISSS}, 2007.

\bibitem{sms}
R.~Ausavarungnirun, K.~Chang, L.~Subramanian, G.~H. Loh, and O.~Mutlu.
\newblock {Staged Memory Scheduling: Achieving high performance and scalability
  in heterogeneous systems}.
\newblock In {\em ISCA}, 2012.

\bibitem{surf}
H.~Bay, A.~Ess, T.~Tuytelaars, and L.~V. Gool.
\newblock {SURF: Speeded Up Robust Features}.
\newblock In {\em CVIU}, 2008.

\bibitem{fst}
E.~Ebrahimi, C.~J. Lee, O.~Mutlu, and Y.~N. Patt.
\newblock {Fairness via Source Throttling: A configurable and high-performance
  fairness substrate for multi-core memory systems}.
\newblock In {\em ASPLOS}, 2010.

\bibitem{weighted-speedup}
S.~Eyerman and L.~Eeckhout.
\newblock {System-level performance metrics for multiprogram workloads}.
\newblock {\em IEEE Micro}, (3), 2008.

\bibitem{resizing}
P.~N. Gour, S.~Narumanchi, S.~Saurav, and S.~Singh.
\newblock Hardware accelerator for real-time image resizing.
\newblock In {\em VLSI Design and Test, 18th International Symposium on}, 2014.

\bibitem{memguard}
Y.~Heechul, Y.~Gang, P.~Rodolfo, C.~Marco, and S.~Lui.
\newblock Memguard: Memory bandwidth reservation system for efficient
  performance isolation in multi-core platforms.
\newblock In {\em RTAS}, 2013.

\bibitem{HWA_sift}
F.-C. Huang, S.-Y. Huang, J.-W. Ker, and Y.-C. Chen.
\newblock {High-Performance SIFT Hardware Accelerator for Real-Time Image
  Feature Extraction}.
\newblock {\em Circuits and Systems for Video Technology, IEEE Transactions
  on}, 22(3), March 2012.

\bibitem{Hur_micro2004}
I.~Hur and C.~Lin.
\newblock {Adaptive History-Based Memory Schedulers}.
\newblock In {\em MICRO}, 2004.

\bibitem{opencv}
Itseez.
\newblock {Open Source Computer Vision}.

\bibitem{jedec-ddr3}
{JEDEC}.
\newblock {Standard No. 79-3. DDR3 SDRAM STANDARD}, 2010.

\bibitem{schedulingCPUGPU}
M.~K. Jeong, M.~Erez, C.~Sudanthi, and N.~Paver.
\newblock {A QoS-aware Memory Controller for Dynamically Balancing GPU and CPU
  Bandwidth Use in an MPSoC}.
\newblock In {\em DAC-49}, 2012.

\bibitem{bank-part}
M.~K. Jeong, D.~H. Yoon, D.~Sunwoo, M.~Sullivan, I.~Lee, and M.~Erez.
\newblock Balancing {DRAM} locality and parallelism in shared memory {CMP}
  systems.
\newblock In {\em HPCA}, 2012.

\bibitem{Hyouseung_rtas2014}
H.~Kim, D.~de~Niz, B.~Andersson, M.~Klein, O.~Mutlu, and R.~R. Roitzsch.
\newblock Bounding memory interference delay in cots-based multi-core systems.
\newblock In {\em RTAS}, 2014.

\bibitem{exnos}
W.~Kim, H.~Chung, H.-D. Cho, and Y.~Kim.
\newblock {Enjoy the Ultimate WQXGA Solution with Exynos 5 Dual}.
\newblock {\em Samsung Electronics White Paper}, 2012.

\bibitem{atlas}
Y.~Kim, D.~Han, O.~Mutlu, and M.~Harchol-Balter.
\newblock {ATLAS}: A scalable and high-performance scheduling algorithm for
  multiple memory controllers.
\newblock In {\em HPCA}, 2010.

\bibitem{tcm}
Y.~Kim, M.~Papamichael, O.~Mutlu, and M.~Harchol-Balter.
\newblock {Thread Cluster Memory Scheduling: Exploiting Differences in Memory
  Access Behavior}.
\newblock In {\em MICRO}, 2010.

\bibitem{Qware}
K.-B. Lee, T.-C. Lin, and C.-W. Jen.
\newblock {An efficient quality-aware memory controller for multimedia platform
  SoC}.
\newblock {\em IEEE Transactions on Circuits and Systems for Video Technology},
  May 2005.

\bibitem{mra}
S.~E. Lee, Y.~Zhang, Z.~Fang, S.~Srinivasan, R.~Iyer, and D.~Newell.
\newblock Accelerating mobile augmented reality on a handheld platform.
\newblock In {\em ICCD}, 2009.

\bibitem{pact-bank-part}
L.~Liu, Z.~Cui, M.~Xing, Y.~Bao, M.~Chen, and C.~Wu.
\newblock A software memory partition approach for eliminating bank-level
  interference in multicore systems.
\newblock In {\em PACT}, 2012.

\bibitem{pin}
C.~K. Luk, R.~Cohn, R.~Muth, H.~Patil, A.~Klauser, G.~Lowney, S.~Wallace, V.~J.
  Reddi, and K.~Hazelwood.
\newblock {Pin: Building customized program analysis tools with dynamic
  instrumentation}.
\newblock In {\em PLDI}, 2005.

\bibitem{beyond}
C.~Macian, S.~Dharmapurikar, and J.~Lockwood.
\newblock Beyond performance: secure and fair memory management for multiple
  systems on a chip.
\newblock In {\em FPT}, 2003.

\bibitem{micron-ddr3}
Micron.
\newblock {1Gb: x4, x8, x16 DDR3 SDRAM Features}.

\bibitem{mcp}
S.~P. Muralidhara, L.~Subramanian, O.~Mutlu, M.~Kandemir, and T.~Moscibroda.
\newblock Reducing memory interference in multicore systems via
  application-aware memory channel partitioning.
\newblock In {\em MICRO}, 2011.

\bibitem{stfm}
O.~Mutlu and T.~Moscibroda.
\newblock Stall-time fair memory access scheduling for chip multiprocessors.
\newblock In {\em MICRO}, 2007.

\bibitem{parbs}
O.~Mutlu and T.~Moscibroda.
\newblock Parallelism-aware batch scheduling: Enhancing both performance and
  fairness of shared {DRAM} systems.
\newblock In {\em ISCA}, 2008.

\bibitem{fqm}
K.~J. Nesbit, N.~Aggarwal, J.~Laudon, and J.~E. Smith.
\newblock Fair queuing memory systems.
\newblock In {\em MICRO}, 2006.

\bibitem{analyzable}
M.~Paolieri, E.~Quiñones, F.~Cazorla, and M.~Valero.
\newblock An analyzable memory controller for hard real-time {CMPs}.
\newblock {\em IEEE Embedded Systems Letters}, Dec 2009.

\bibitem{pinpoint}
H.~Patil, R.~Cohn, M.~Charney, R.~Kapoor, A.~Sun, and A.~Karunanidhi.
\newblock Pinpointing representative portions of large {Intel} {Itanium}
  programs with dynamic instrumentation.
\newblock In {\em MICRO}, 2004.

\bibitem{snapdragon}
Qualcomm.
\newblock {Snapdragon S4 Processors: System on Chip Solutions for a New Mobile
  Age}.
\newblock {\em Qualcomm White Paper}, 2011.

\bibitem{PRET}
J.~Reineke, I.~Liu, H.~D. Patel, S.~Kim, and E.~A. Lee.
\newblock {PRET DRAM} controller: Bank privatization for predictability and
  temporal isolation.
\newblock In {\em CODES+ISSS}, 2011.

\bibitem{frfcfs}
S.~Rixner, W.~J. Dally, U.~J. Kapasi, P.~Mattson, and J.~D. Owens.
\newblock Memory access scheduling.
\newblock In {\em ISCA-27}, 2000.

\bibitem{HWA_acoustic}
I.~Schmadecke and H.~Blume.
\newblock Hardware-accelerator design for energy-efficient acoustic feature
  extraction.
\newblock In {\em Consumer Electronics (GCCE), 2013 IEEE 2nd Global Conference
  on}, 2013.

\bibitem{ws}
A.~Snavely and D.~M. Tullsen.
\newblock Symbiotic jobscheduling for a simultaneous multithreaded processor.
\newblock In {\em ASPLOS}, 2000.

\bibitem{Sobel}
I.~Sobel.
\newblock An isotropic 3x3 image gradient operator.
\newblock In {\em Machine Vision for Three-Dimensional Scenes}, pages 376--379.
  Academic Press, 1990.

\bibitem{mobileeye}
G.~P. Stein, I.~Gat, and G.~Hayon.
\newblock {Challenges and Solutions for Bundling Multiple DAS Applications on a
  Single Hardware platform.}
\newblock In {\em V.I.S.I.O.N.}, 2008.

\bibitem{armwhite}
A.~Stevens.
\newblock {{QoS for High-Performance and Power-Efficient HD Multimedia}}.
\newblock {\em ARM White Paper}, 2010.

\bibitem{blist}
L.~Subramanian, D.~Lee, V.~Seshadri, H.~Rastogi, and O.~Mutlu.
\newblock The blacklisting memory scheduler: Achieving high performance and
  fairness at low cost.
\newblock In {\em ICCD}, 2014.

\bibitem{mise}
L.~Subramanian, V.~Seshadri, Y.~Kim, B.~Jaiyen, and O.~Mutlu.
\newblock {MISE}: Providing performance predictability and improving fairness
  in shared main memory systems.
\newblock In {\em HPCA}, 2013.

\bibitem{visconti2}
Y.~Tanabe, M.~Sumiyoshi, M.~Nishiyama, I.~Yamazaki, S.~Fujii, K.~Kimura,
  T.~Aoyama, M.~Banno, H.~Hayashi, and T.~Miyamori.
\newblock A 464{GOPS} 620{GOPS/W} heterogeneous multi-core {SoC} for
  image-recognition applications.
\newblock In {\em ISSCC}, 2012.

\bibitem{max-slowdown}
H.~Vandierendonck and A.~Seznec.
\newblock Fairness metrics for multi-threaded processors.
\newblock {\em IEEE CAL}, February 2011.

\bibitem{haarlike}
P.~Viola and M.~Jones.
\newblock Rapid object detection using a boosted cascade of simple features.
\newblock In {\em CVPR}, 2001.

\bibitem{timepred}
L.~Wu and W.~Zhang.
\newblock Time-predictable {DRAM} access scheduling algorithms for real-time
  multicore processors.
\newblock In {\em Southeastcon}, 2013.

\end{thebibliography}
}

\end{document}